\documentclass[12pt]{iopart}
\usepackage{iopams}
\usepackage{url}
\usepackage{graphicx,color}
\usepackage{verbatim}
\usepackage{textalpha}

\usepackage[LGR,T1]{fontenc}
\usepackage[utf8]{inputenc}

\usepackage{hyperref}

\graphicspath{{./figs/}}

\newcommand{\binom}[2]{{#1 \choose #2}}
\newcommand{\Nuc}[1]{\mathrm{#1}}

\begin{document}
\title{Mutation supply and the repeatability of selection for antibiotic resistance}
\author{Thomas van Dijk$^{1}$\footnote{These authors contributed equally.}, Sungmin Hwang$^{2}$$ \ddag $, Joachim Krug$^2$, J. Arjan G.M. de Visser$^1$, Mark P. Zwart$^{1,2}$\footnote{Corresponding author:} }
\address{$^1$Laboratory of Genetics, Wageningen University, Wageningen, The Netherlands}
\address{$^2$Institute of Theoretical Physics, University of Cologne, Cologne, Germany}
\vspace{10pt}
\begin{indented}
\item[]\today
\end{indented}
\begin{abstract}
Whether evolution can be predicted is a key question in evolutionary biology. Here we set out to better understand the repeatability of evolution, which is a necessary condition for predictability. We explored experimentally the effect of mutation supply and the strength of selective pressure on the repeatability of selection from standing genetic variation. Different sizes of mutant libraries of an antibiotic resistance gene, TEM-1 β-lactamase in \textit{Escherichia coli}, were subjected to different antibiotic concentrations. We determined whether populations went extinct or survived, and sequenced the TEM gene of the surviving populations. The distribution of mutations per allele in our mutant libraries\textemdash generated by error-prone PCR\textemdash followed a Poisson distribution. Extinction patterns could be explained by a simple stochastic model that assumed the sampling of beneficial mutations was key for survival. In most surviving populations, alleles containing at least one known large-effect beneficial mutation were present. These genotype data also support a model which only invokes sampling effects to describe the occurrence of alleles containing large-effect driver mutations. Hence, evolution is largely predictable given cursory knowledge of mutational fitness effects, the mutation rate and population size. There were no clear trends in the repeatability of selected mutants when we considered all mutations present. However, when only known large-effect mutations were considered, the outcome of selection is less repeatable for large libraries, in contrast to expectations. Furthermore, we  show experimentally that alleles carrying multiple mutations selected from large libraries confer higher resistance levels relative to alleles with only a known large-effect mutation, suggesting that the scarcity of high-resistance alleles carrying multiple mutations may contribute to the decrease in repeatability at large library sizes. 


\end{abstract}
%
\submitto{\PB}
%

\section{Introduction}
One issue that has taken center stage in evolutionary biology in recent years is the question of
whether evolution is predictable \cite{Visser2014}.
Various factors have driven this question to the foreground.
First, one stringent test of a scientific framework is its ability to make quantitative
predictions about future events (E.g., \cite{Dyson1920}). Experimental evolution \cite{Buckling2009}, but also new approaches
in paleontology \cite{Daeschler2006}, have increased the scope for hypothesis testing in evolutionary biology.
Making clear predictions about future evolution, even for simple laboratory model systems at short time scales,
would demonstrate a profound understanding of evolutionary processes.
However, the extent to which non-trivial predictions actually can be made for such a complex and inherently stochastic process is unclear.
Second, the widespread occurrence of convergent phenotypic evolution raises the question of what drives these parallel changes in natural populations.
Moreover, recent observations suggest that the genetic basis for these adaptations is often divergent \cite{Soria-Carrasco2014,Yeaman2016,Natarajan2016}, making
this question even more intriguing. Finally, the predictability of evolution has important practical implications, such as whether future changes in pathogen populations can be anticipated \cite{Luksza2014} and antibiotic resistance can be avoided or mitigated \cite{Barlow2002,MacLean2010}.

One important concept that has motivated these developments is the fitness landscape, a
multi-dimensional model that maps different genotypes to their associated fitness \cite{Wright1932}. The topography of fitness landscapes is strongly contingent on epistatic interactions among loci,
particularly sign epistasis.
Sign epistasis occurs when mutations can either be beneficial or
deleterious, depending on the background in which they appear \cite{Weinreich2006,Poelwijk2007}.
Sign epistasis has two ramifications for the predictability of evolution.
First, it constrains the number of different pathways accessible to natural selection, thereby potentially contributing to a greater predictability of evolution.
Second, it can shape the fitness landscape in such a way that it contains multiple peaks, and thus multiple distinct end-points for evolution \cite{Poelwijk2011} and thereby a
lower predictability. Weinreich and colleagues demonstrated the pervasiveness of sign epistasis
in a complete fitness landscape for five mutations in the TEM-1 β-lactamase gene \cite{Barlow2002,Salverda2010}. Of the
120 possible pathways by which these mutations could be acquired, under strong selection/weak mutation conditions the expectation is that only 18 pathways will be accessible and that most populations will follow 2-4 pathways \cite{Weinreich2006}.
Subsequent research demonstrated that sign epistasis occurs between two large-effect initial mutations, R164S and
G238S, and that these initial mutations lead to separate adaptive trajectories \cite{Salverda2011,Schenk2013,Dellus-Gur2015}. The
dependence of evolutionary trajectories on historical contingencies also has been shown in other
model systems \cite{Blount2012,Tenaillon2012,Kryazhimskiy2014}, as well as in natural populations \cite{Natarajan2016}.
Populations can follow different mutational pathways because of the combined effects of genetic drift and epistasis, and therefore not necessarily because of the ruggedness of the fitness landscape \cite{Kryazhimskiy2014}.

Another key factor for predicting evolution is population size, which will strongly affect
the repeatability of evolution.
The repeatability of a process will determine the extent to which it can be predicted. 
Genetic drift will be stronger in small populations, and the mutational supply will be smaller.
If the mutational supply is small, populations are likely to either sample no beneficial mutations, or to fix whichever beneficial mutation is present.
Hence, independent populations are likely to follow different evolutionary trajectories.
Conversely, in larger populations multiple beneficial mutations can be sampled simultaneously
in different lineages. However, in asexual populations, a beneficial combination of mutations
cannot be obtained through recombination, but only when these beneficial mutations occur simultaneously
or sequentially in the same lineage. The largest effect mutation will therefore likely prevail, whilst taking longer to fix because there is competition between beneficial
mutations leading to small effective selection coefficients among contending mutants. This effect is called “clonal interference” \cite{Fisher1930,Muller1932,Crow1965,Gerrish1998}, and it will limit the number of mutations that can be fixed and hereby increase
repeatability. Indeed, experimental evolution studies have shown that the repeatability of evolution increases with population size \cite{Rozen2008,Lachapelle2015}.

On the other hand, recent studies have shown that the relationship between population
size and the repeatability of evolution can be more complex. Theoretical work has shown
that the effects of population size on the repeatability of evolution can be non-monotonic: for
very large populations adapting on a rugged fitness landscape the repeatability of evolution can decrease \cite{Szendro2013,Ochs2015}. These large
populations will sometimes sample rare, highly beneficial combinations of multiple mutations that would otherwise be evolutionarily inaccessible due to epistasis
\cite{Szendro2013}. Genetic drift will also allow valley crossing in some small populations, whereas intermediate-size populations will instead greedily converge on a local fitness peak most of the time \cite{Rozen2008,Ochs2015}. Examples of non-monotonic relations between population size and evolutionary predictability have also been observed experimentally, although they implicate a different underlying mechanism. For \textit{Pseudomonas fluorescens} adaptation to rifampicin, the greatest diversity in the genetic basis of adaptation occurred at intermediate population sizes \cite{Vogwill2016}. In this case survival of populations at the smallest bottleneck size was contingent upon the occurrence of large-effect mutations: those mutations that confer growth rates large enough to overcome the effect of the larger dilutions imposed on these populations. 

The fitness landscape and population size will be important
determinants of the repeatability of evolution, but other more subtle,
system-specific factors may also come into play. Natural selection
is conceptually a purely deterministic force in evolution, as opposed
to the inherently stochastic processes of mutation and genetic drift
\cite{Fisher1930}, and therefore a key driver of evolutionary
predictability. Furthermore, whereas the importance of directional and
purifying selection for evolutionary convergence may be obvious, even
diversifying selection can lead to highly repeatable
evolutionary outcomes \cite{Herron2013}. However, non-heritable phenotypic variation
will affect the outcome of selection and therefore the extent to which
the selection of genotypes can be a deterministic process. One extreme
example of non-heritable phenotypic variation is the bistable rate of growth in
bacterial populations, which probably represents a  bet hedging strategy \cite{Deris2013}. Moreover, many other mechanisms could
impact stochasticity in natural selection by amplifying the effects of
minor stochastic variation during selection. Social interactions
between helpers and beneficiaries \cite{Yurtsev2013} could have such effects; If a helper is at low frequency and there is variation in its growth rate, this could affect the breakdown of an antibiotic and thereby the scope for growth of the beneficiary. Furthermore, in chaotic systems, such as populations with beneficiaries that interfere with the growth of helpers \cite{Kirkwood1994,Zwart2013}, small perturbations caused by selection can have major effects on the outcome of evolution.

Here we explore the relationship between mutation supply and the repeatability of evolution in a simple and well characterized experimental model system. As a first step, we wanted to consider the effects of mutation supply on the repeatability of selection, using a range of mutation supplies and selection pressures. Specifically, we were interested in determining the contribution of mutation and selection to the outcome of a single round of evolution, in a controlled laboratory environment. Can stochasticity inherent to selection have important implications for the outcome of evolution? Second, we were interested in whether a limited number of large-effect mutations can dominate selection, despite epistatic interactions with other mutations. To what extent does the occurrence of large-effect mutations shape the outcome of selection? We choose to study these questions with mutagenized libraries of TEM-1 β-lactamase in \textit{E. coli}, and used the antibiotic cefotaxime to introduce a selection pressure \cite{Barlow2002}. This model system is well suited to addressing these issues because the type and intensity of selection can be tuned by using different cefotaxime concentrations. Moreover, because mutations are induced by polymerase chain reaction (PCR) and the mutation supply can be further regulated by making libraries of different sizes, we can manipulate the main evolutionary forces independently and tailor a setup to specifically consider stochastic effects during selection. Finally, the repeatability and predictability of evolution have been extensively studied for this model system, helping to provide substantial context for this work.

\section{Results}
\subsection{Experimental approach}

To investigate the effects of standing genetic diversity and selection pressure on the repeatability of selection, we generated libraries of TEM β-lactamase alleles by error-prone PCR (EP-PCR) and cloned them into a plasmid vector. We refer to the ensemble of TEM alleles generated by a single EP-PCR as a “pool”. From these pools we then generated "libraries" with a range of determined sizes. To do so, we estimated the number of transformed cells, and then diluted the cells to render four different-sized “libraries” containing $ 10^2, 10^3, 10^4$ and $10^5$ TEM alleles. These cells were then amplified by overnight culturing, and subsequently exposed to four different concentrations of cefotaxime (64, 128, 256 and 512 ng/mL cefotaxime). To avoid biases related to initial cellular density \cite{Yurtsev2013, Artemova2015}  and associated differences in cell doublings during selection, the same density ($1.1\times 10^5$ cells/mL) was used for all treatments.  As $10^6$ cells were used to inoculate each replicate, even for the largest library size ($10^5$) the predicted number of alleles lost during inoculation is small, but larger than in small libraries. Note that the lowest antibiotic concentration used is just above that in which \textit{E. coli} carrying the original TEM-1 gene will grow: The minimum inhibitory concentration (MIC) is $\approx 60$ ng/mL \cite{Salverda2011}. Each of the combinations of population size and antibiotic concentration was tested with six-fold  replication. We refer to these replicates as "physical replicates" (analogous to technical replicates in the vernacular of biologists). The entire experiment was carried out with libraries drawn from three independently generated EP-PCR pools (refered to as Pools A, B and C), and these we refer to as "biological replicates". Following selection, the consensus sequence of the TEM gene was determined by Sanger sequencing. We characterized the mutant libraries, analyzed the extinction rates with a simple model, and determined the diversity of TEM alleles present in the selected populations to assess repeatability.

\subsection{Statistical description of error-prone PCR libraries}
\label{sec:StatisticsLibs}
To know the mean and distribution of single-nucleotide mutations in the EP-PCR pools, we sequenced six clones from each pool. In these 18 clones, we detected 27 single-nucleotide mutations and hence the mean number of single-nucleotide mutations per allele ($ \lambda $) is 1.5. A one-sample Kolmogorov-Smirnov test confirmed that the distribution of the number of single-nucleotide mutations per allele is not significantly different than that of a Poisson distribution with a mean of 1.5 ($ D = 0.232, N = 18,  P = 1 $, see \Fref{fig:EPLibraries}a). For the three pools, $ \lambda $ was similar suggesting the distributions are comparable ($ \lambda_A = 1.33, \lambda_B = 1.66, \lambda_C = 1.5 $). Our observed distribution of mutations per allele matches theoretical expectations from a simple model \cite{Sun1995}, but clashes with empirical data with significantly more variation than a Poisson distribution \cite{Drummond2005}. We have introduced far fewer mutations per amplicon here (1.5 here vs. 15.8 and 19.9 in \cite{Drummond2005}), probably making our setup less susceptible to subtle deviations from the Poisson prediction. The position of all 27 mutations detected in the libraries (\Fref{fig:EPLibraries}b) was found to be not significantly different from a uniform distribution using a one-sample Kolmogorov-Smirnov test ($ D = 0.672, N = 27, P = 0.758 $; range is 1-861, i.e., the length of the TEM ORF). This indicates that mutations are randomly distributed over the TEM gene. The mutational spectrum for the 27 mutations was found to be similar to expectations for the GeneMorph II random mutagenesis kit \cite{Anonymous2015}, with an overrepresentation of transitions and t $\to$ a transversions (\Fref{fig:EPLibraries}c).

Given these results, it is reasonable to assume the distribution of mutations per TEM allele follows a Poisson distribution for our setup, whilst the mutational biases of the EP-PCR should probably be taken into account by a model. It also should be kept in mind that mutations that occur early during the PCR will reach higher frequencies and some coordination of mutations within a pool, as well as the libraries derived from it, will occur \cite{Sun1995}. Therefore, alleles drawn from the same pool are more likely to share specific mutations than alleles drawn from different pools.

\subsection{Extinction rates during the selection experiment}
\label{sec:ExtinctionRates}
\begin{figure*}[t]
	\centering
	\includegraphics[width=0.9\textwidth]{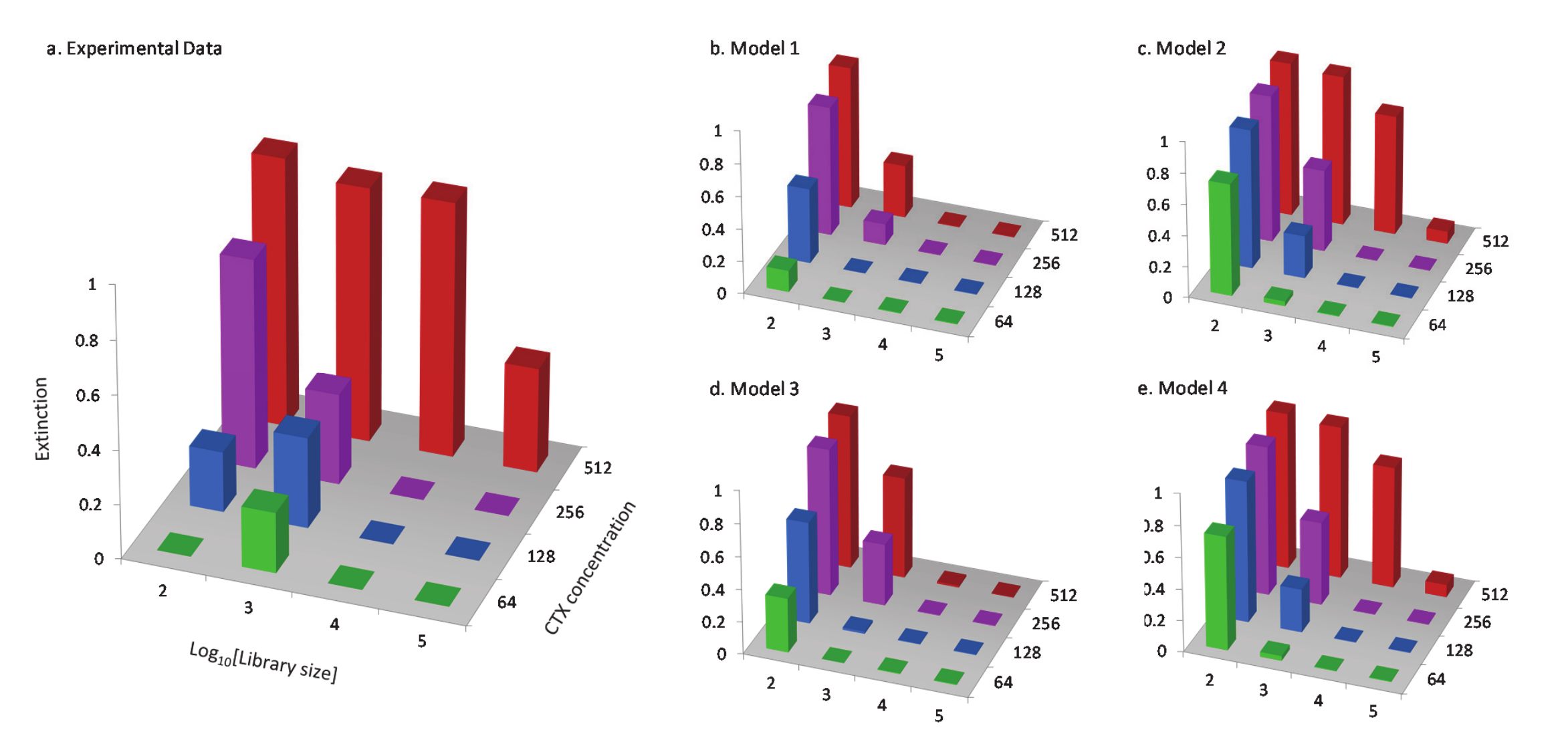}
	\caption{
		The observed and modeled rates of extinction for the various library sizes exposed to different concentrations of cefotaxime are shown. In panel a, the experimental data are given, extinction being the mean of the three experimental replciates. In panels b to e, extinction for the fitted Models 1-4 is given (see also \Tref{tab:ExtinctionModelPar}).
	}
	\label{fig:ExitinctionModel}
\end{figure*}

Selection experiments were performed for various library sizes and over a range of cefotaxime concentrations. Overall, the percentage of populations in which bacterial growth could not be detected after 48 h was $ 32.3 \pm 8.3 $. We subsequently refer to populations without detectable growth as having gone extinct. The rate of extinction appears to increase for smaller library sizes and higher cefotaxime concentrations (\Fref{fig:ExitinctionModel}a). We used logistic regression to test these trends, where library size was nested within EP-PCR pool due to the coordination of mutations. The effects of pool, library size within pool, and cefotaxime concentration within library size within pool were all highly significant ($ P < 0.001 $).

We generated a simple model of extinction to test whether the observed patterns of extinction can be explained, and are compatible with previous results. A brief overview is given here, and a more detailed description of the model and model fitting routine are provided in the methods section.  In essence, for a library of a given size the model draws a Poisson-distributed number of mutations for each allele, and determines whether the mutations are beneficial or not. For the beneficial mutations, the model assigns resistance values to each individual mutation based on an empirical distribution of beneficial fitness effects (DBFE) for TEM-1 \cite{Schenk2012}, weighted to account for mutational biases of the polymerase. The resistance of each allele is then determined by assuming the effects of multiple mutations are either additive, or alternatively, that each (additional) mutation has an epistatic effect of a constant magnitude $ \epsilon $. The DBFE for TEM-1 is based on IC$ _{99.99} $ measurements \cite{Schenk2012}, and since these measurements are based on colony counts, they may not accurately reflect resistance under our selection conditions. We therefore also introduced a factor $ \rho $ that allows for the rescaling of all resistance values. We then draw $ L $ TEM alleles, with $ L $ corresponding to library size, and impose hard selection: If any of these alleles has a higher resistance then the antibiotic concentration being considered, the population grows and does not go extinct.

We then considered four variants of this extinction model. Model 1 assumes no epistasis ($ \epsilon = 0 $) and no rescaling of resistance ($ \rho = 0 $). Model 2 allows only rescaling of resistance, Model 3 allows only epistasis and Model 4 allows both rescaling of resistance and epistasis. Model 1 has no free parameters and does not need to be fitted, whereas Models 2-4 were fitted to the empirical extinction data using a maximum likelihood approach, and the Akaike information criterion (AIC) was used for model selection. Models 1 and 3 are not supported by the data, Models 2 and 4 have similar fits, and there is slightly more support for Model 2 given it has one less free parameter (\Tref{tab:ExtinctionModelPar}, \Fref{fig:ExitinctionModel}b-e). This suggests that any epistatic effects are weak ($ \rho = -0.1 $), and that rescaling resistance values of the DBFE is the most important consideration. The observed rates of extinction are therefore compatible with model predictions, suggesting the absence of beneficial mutations in TEM drive extinction in small populations and at high antibiotic concentrations. Moreover, these results suggest that epistasis does not play a major role in determining resistance levels in this setup. However, we are cautious with drawing this conclusion because the limited scope for interactions between mutations, due to the low number of mutations per allele in the pools used here. Moreover, if positive and negative epistatic effects balance out on average, modeling epistatic effects as being constant would not be illuminating.

\begin{table*}
\centering
\caption{Model parameter estimates and model selection. $ ^* $ : Akaike Weight}
	\begin{tabular}{|c|c|c|c|c|c|c|}
		\hline 
		& \multicolumn{2}{c|}{Parameter estimates} &  &  &  &  \\ 
		\hline 
		Model & $\rho$ & $\epsilon$ & NLL & AIC & $\Delta$AIC & AW$ ^{*} $ \\ 
		\hline 
		1 & - & - & 197.491 & 394.982 & 243.862 & 0 \\ 
		\hline 
		2 & -1.2 & - & 74.560 & 151.121 & - & 0.624 \\ 
		\hline 
		3 & - & -2.8 & 151.592 & 305.148 & 154.063 & 0 \\ 
		\hline 
		4 & -1.1 & -0.1 & 74.067 & 152.134 & 1.103 & 0.376 \\ 
		\hline 	
	\end{tabular} 
	\label{tab:ExtinctionModelPar}
\end{table*}

\subsection{TEM genotypes in selected populations}
\begin{figure*}[t]
	\centering
	\includegraphics[width=0.9\textwidth]{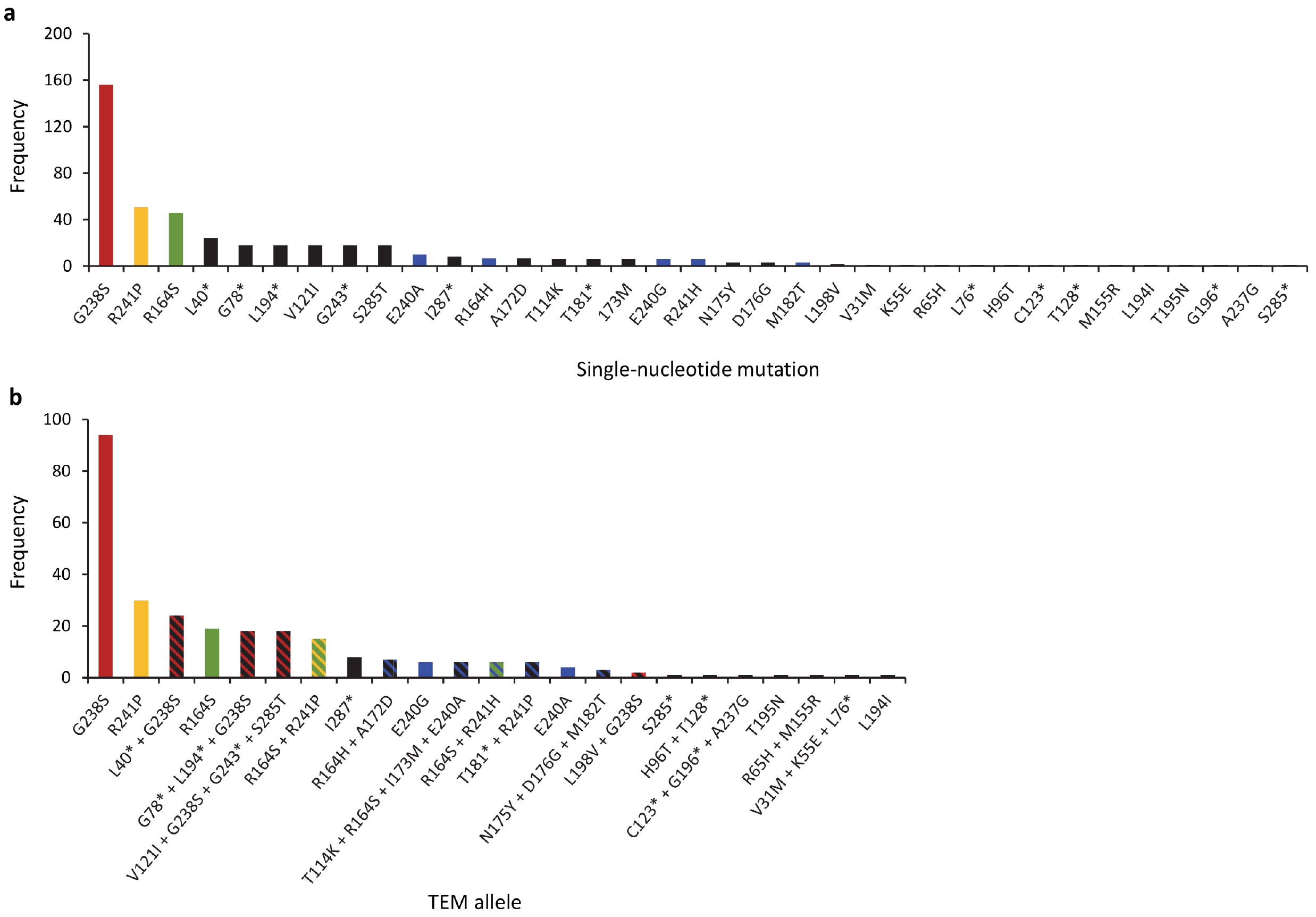}
	\caption{
In panel a, the frequency of single-nucleotide mutations detected in all selected populations is given. In panel b, the frequency at which TEM alleles were observed is given. For both panels, only mutations or alleles detected in more than one population are shown. Unknown mutations, or alleles with only unknown mutations, are colored solid black. Known mutations, or alleles containing only one known mutation, are given a solid color, with the less prevalent mutations all colored in light blue. All known mutations are beneficial on their own, with the exception of M182T, which is beneficial only in combination with other mutations. In panel b, alleles containing combinations of mutations are given hatched colors corresponding to the mutations present. Whereas a considerable number of unknown mutations were detected, most alleles contain at least one known beneficial mutation. The mutation G238S dominates, as do alleles containing this mutation alone or in combination with other mutations. The ancestral TEM-1 allele was inferred to be present in 38 populations, although its presence could represent a consortium of mutant alleles at low frequencies.  
	}
	\label{fig:MutationsAlleles}
\end{figure*}

We detected a total of 453 single-nucleotide mutations in the populations following selection, with an average of 2.32 mutations per surviving population. The most abundant mutation was G238S, the largest effect single-nucleotide mutation in TEM-1 \cite{Schenk2012} (\Fref{fig:MutationsAlleles}a). The three most common mutations (G238S, R241P and R164S) had all been previously identified, and are among the four largest-effect mutations known for TEM-1 adaptation to cefotaxime \cite{Schenk2012}. Overall, seven out of 35 detected mutations are known beneficial mutations \cite{Schenk2012} (\Fref{fig:MutationsAlleles}a), and the stabilizing mutation M182T was also observed \cite{Sideraki2000}. Of the detected mutations, 24\% were synonymous. Ten synonymous mutations that increase resistance to cefotaxime have been identified in TEM-1 \cite{Schenk2012}, but none of these mutations were found here.

We inferred the presence of 273 TEM alleles carrying mutations in the surviving populations, representing 23 different alleles (\Fref{fig:MutationsAlleles}b, \Fref{fig:Library1}-\ref{fig:Library3}). The presence of the ancestral TEM-1 allele was also inferred in many selected populations, although its presence could also be due to a consortium of many different variants at low frequencies. When we sequenced clones to check that the TEM alleles inferred were actually present in the populations, we indeed found that there were sometimes additional mutations present which could not be seen at the population level. We found an average of 1.40 and a maximum of five TEM alleles per surviving population. We analyzed the occurrence of mixed-allele populations with a generalized linear mixed model with a binomially distributed response variable, with EP-PCR pool as a random factor, and  as fixed factors library size within pool and cefotaxime concentration within library size within pool. The latter two factors were highly significant ($ P < 0.001 $), whereas pool was not ($ P > 0.05 $). The model coefficients show that the frequency of mixed-allele infections increased with library size ($ 0.755 \pm0.151 $), but decreased with cefotaxime concentration within library size ($ -0.234 \pm0.047 $). Therefore, mixed-allele populations are most common in large libraries and at low cefotaxime concentrations, as one would expect a priori given the occurrence of hard selection.

Whereas approximately one fifth of the mutations
detected were previously identified beneficial mutations
(\Fref{fig:MutationsAlleles}a), almost all TEM alleles contained at
least one known beneficial mutation
(\Fref{fig:MutationsAlleles}b). Many of the unknown mutations in the
selected haplotypes therefore may be hitchhiking with previously
identified beneficial mutations. One interesting exception might be
the synonymous mutation I287*, which was found on its own in eight
haplotypes. Another exception is the L40* mutation, which occurred
together with G238S in the haplotype that dominated the largest size
library from EP-PCR pool B (\Fref{fig:Library2}).  Interestingly, this L40* and G238S double mutant consistently grew at 512 ng/ml cefotaxime, whereas G238S on its own could not grow at this concentration under our assay conditions (\Fref{fig:Library1}-\ref{fig:Library3}). These observations suggest L40* is beneficial, when in combination with G238S. 

One striking feature in the selected populations is how similar or variable the outcome of selection can be. In many cases, the same allele is fixed in all six replicate populations at one antibiotic concentration, or even in all replicate populations at all antibiotic concentrations of a given library size (E.g., \Fref{fig:Library2}, library sizes $10^3$ and $10^5$). These dominant alleles are probably generalists that possess both high resistance and competitive ability in the absence of antibiotic. On the other hand, at low antibiotic concentrations the outcome of selection appears to be quite variable. For example, for the library size 100 of pool C (\Fref{fig:Library3}), no known large-effect mutations are observed after selection. The resulting population therefore depends on competition between those alleles that survive initial selection, and is therefore highly variable, especially at 128 ng/mL cefotaxime.

Conversely, in some of the largest library sizes the outcome of selection can also be variable. One interesting case is the $10^5$ library size of pool C, in which three different TEM alleles are present in the surviving populations. These alleles all contain either R164S, R164H or G238S. The initial fixation of either R164S or G238S commits a population to distinct evolutionary trajectories \cite{Salverda2011,Schenk2013,Dellus-Gur2015}. Therefore, in this example the variability in the outcome of selection would force these populations down two different trajectories in subsequent evolution. A previous study   already argued that for selection on large EP-PCR libraries, the fixation of R164S or G238S was not due to the absence of either mutation, but rather the combined effects of clonal interference and epistasis on selection \cite{Salverda2011}. Our observations here show that random effects during the selection phase are strong enough to variably commit a population to either trajectory, highlighting the importance of these random effects in limiting the repeatability of evolution.

\subsection{Resistance and fitness of G238S-containing alleles}
In the selected populations, we noticed that TEM alleles containing only the G238S mutation appear to fix for libraries of size $10^4$ from all three pools, at the highest cefotaxime concentration with growth, 256 ng/mL (see \Fref{fig:Library1}-\ref{fig:Library3}). Sanger sequencing of clones confirmed that these populations consisted almost exclusively of alleles with only the G238S mutation, and not an ensemble of alleles containing G238S and other mutations. However, in selected populations from the $10^3$ and $10^5$ size libraries, alleles containing G238S together with other mutations are fixed. In the large libraries we expect that the additional mutations might further increase resistance and fitness, as these large libraries will sample more variation from the EP-PCR pool and because these alleles grow at a higher cefotaxime concentration than that at which G238S alone is observed (512 ng/mL). On the other hand, for the smaller libraries we postulated that these additional mutations might be neutral or even deleterious, since any allele containing G238S is likely to have a high enough resistance to dominate such small populations.

To test these hypotheses, we measured the resistance (MIC) and relative fitness (\textit{W}) of two alleles fixed in the $10^3$ size libraries (S1 and S2) and two from the $10^5$ size libraries (L1 and L2). All alleles except S1 had a significantly higher resistance (\Tref{tab:Resistance}) and relative fitness (\Fref{fig:Fitness}) than the G238S allele. Therefore, our hypothesis that the alleles sampled in $10^5$ size libraries contained additional beneficial mutations was correct, but in one case such an allele was also sampled in a smaller $10^3$ library. Hence, we conclude that the fixation of G238S at an intermediate library size of $10^4$ is probably just a chance occurrence, made possible because alleles with G238S and additional beneficial mutations were not sampled.

Two of the three alleles with higher resistance tested contained a synonymous mutation, and allele L1 contains only the synonymous mutation A184* in addition to G238S (\Tref{tab:Resistance}). These observations are therefore direct evidence that synonymous mutations in TEM-1 β-lactamase can be beneficial for resistance. Our results \textemdash including the repeated occurrence of other synonymous mutations in the selected populations \textemdash therefore support the idea that synonymous mutations in TEM-1 can be beneficial for resistance against cefotaxime, in agreement with previous results \cite{Schenk2012,Firnberg2014}.


\begin{table*}
	\centering
	\caption{MIC values for selected G238S alleles. $^a$: Library size from which the allele originates. $^b$: Minimal inhibitory concentration in $\mu$g/mL cefotaxime. $^c$: Test values for U-test against G238S. $^* $: Statistical significant at the 0.01 level.}
	\begin{tabular}{|c|c|c|c|c|c|c|}
	\hline 
Allele name & Library size$^a$ & Mutations & MIC$^b$ & Z value$^c$ & P value$^c$ \\ 
\hline 
G238S &  & G238S & 2 &  &  \\ 
\hline 
S1 & $10^3$ & G78*, L194*, G238S & 4 & -2.211 & 0.065 \\ 
\hline 
S2 & $10^3$ & V121H, G243*, G238S, S285T & 4 & -2.762 & 0.009* \\ 
\hline 
L1 & $10^5$ & A184*, G238S & 8 & -2.900 & 0.004* \\ 
\hline 
L2 & $10^5$ & L198V, G238S & 8 & -2.866 & 0.004* \\ 
\hline 	
	\end{tabular} 
	\label{tab:Resistance}
\end{table*}

\subsection{Repeatability of selection}
In order to analyze the repeatability of selection in our experiments,
we consider Jaccard distance measures between pairs of allele frequencies of the surviving populations. For the purpose of
this analysis, the haplotype frequencies shown in
\Fref{fig:Library1}-\ref{fig:Library3} were reduced to single-nucleotide mutation frequencies by
assigning to each mutation the sum over the frequencies of haplotypes
in which it is contained. This results in a set of (unnormalized)
allele frequencies $p_i$ where the index $i$ runs over the set $S_0$ containing
all single-nucleotide mutations that appeared at sufficient frequency (> 0.1) to
be included in the haplotype reconstruction, as well as a separate
allele representing the unmutated TEM-1. Finally, the $p_i$ are
rescaled to ensure normalization $\sum_{i \in S_0} p_i =1$.  
Given two sets of allele frequencies $\{p_i\}_{i \in S_0}$ and
$\{q_i\}_{i \in S_0}$, the Jaccard distance is then calculated from the equation 
\begin{eqnarray}
	J(p,q) = 1 - \frac{\sum_i  \min(p_i, q_i)}{\sum_i  \max(p_i, q_i)}, 
\end{eqnarray}
which measures the dissimilarity between the two sets in terms of a
real value in the range of $0 \le J(p,q) \le 1$.

By construction, this measure cannot be applied to populations that have gone extinct because no normalized allele frequencies can be assigned to them.
To circumvent this problem, we simply consider the Jaccard distance of populations conditioned on survival. 
Alternatively, we might have assigned a special state to extinct
populations and treated it as either monomorphic or as maximally
polymorphic. 
However, we found that the effect of such variations on the average
Jaccard distance is not significant, as the trends in the data are nonetheless driven by the allele frequencies of the surviving populations.
Therefore, we restrict ourselves to the simplest scenario in which the populations that have gone extinct are excluded.
In this regard, the following analysis is complementary to the
extinction model presented (Section \ref{sec:ExtinctionRates}) and
focuses on different information contained in the end-point populations.

For a fixed condition distinguished by the library size and cefotaxime
concentration, a total of $3 \times 6 = 18$ populations corresponding
to the 3 independent EP-PCR pools and 6 physical replicates were created. 
Eliminating the physical replicates for the extinct populations, at
most 18 samples are thus available for each condition. 
The Jaccard distances for all pairs of these populations are then
calculated and subsequently classified into two categories,
within-group and between-group distances, based on the EP-PCR pool A, B, and C from which they were prepared.
The averages over different combinations of EP-PCR pools, i.e.,
over A, B and C for within-group Jaccard distance and over pairs A-B,
B-C, C-A for between-group Jaccard distances, provide measures of repeatability on two different levels.
By comparing the diversity measures between the two categories, we can
estimate the relative effects of the two main sources of randomness in
our system, i) the stochasticity of the EP-PCR process and ii) the stochasticity of the selection experiments in the presence of antibiotics, respectively.
Borrowing the well-known concepts of diversity from the field of
ecology, one may regard the within-group Jaccard distance as analogous
to the alpha diversity, which measures the diversity on local scales,
whereas the between-group Jaccard distance is a proxy for the gamma
diversity that quantifies the overall diversity in the entire system \cite{Whittaker1960}.

We then repeated the same procedure by focusing only on subsets of
alleles corresponding to ``driver'' mutations,
i.e. known substitutions of substantial resistance effect.
In order to make the comparison fair, after deleting the frequencies
of non-driver mutations the allele frequencies  were renormalized to one. 
If no driver mutations are present in a surviving population, these
populations are treated as if they went extinct; only a handful of
such cases were observed at the lowest concentrations of cefotaxime. 
By following this approach, we expect the effects of hitchhiking
random mutations and other sources of noise to be reduced.
As possible collections of driver mutations, we introduce two allele
sets $S_1= \{\mathrm{G238S}, \mathrm{R241P}, \mathrm{R164S},
\mathrm{R164H}, \mathrm{E240G}, \mathrm{R241H}\} $ and $ S_2=
\{\mathrm{G238S}, \mathrm{R241P}, \mathrm{R164S}\} $ by reading off
the six and three largest effect single-nucleotide mutations from the
list of known mutations \cite{Schenk2012} present in the selected populations described here, respectively.

\begin{figure*}[t]
	\centering
	\includegraphics[width=0.9\textwidth]{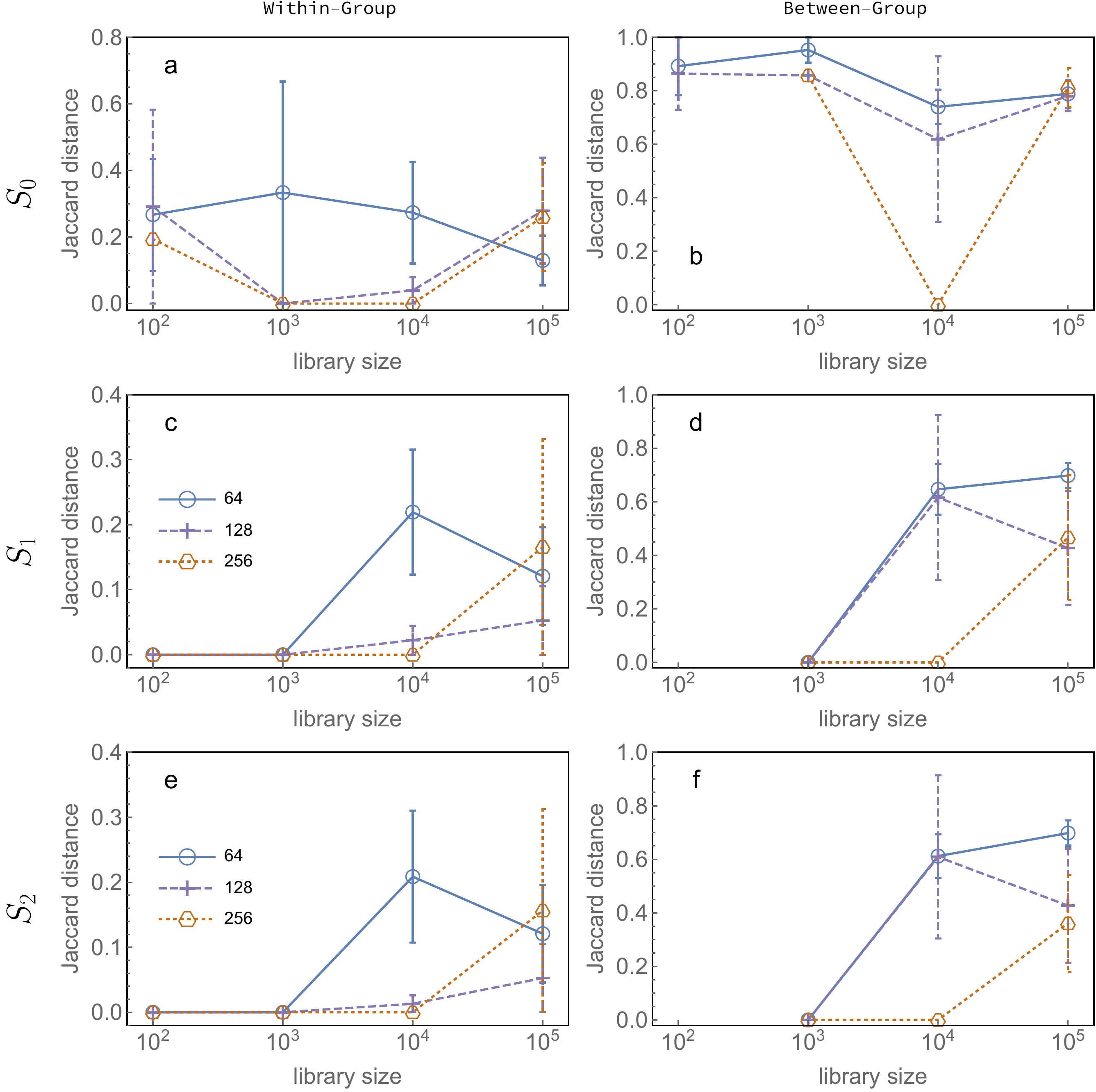}
	\caption{
		Panels show the average within-group (a,c,e) and
                between-group (b,d,f) Jaccard distances of allele
                frequencies in populations exposed to different
                concentrations of cefotaxime as a function of library
                size. In the top row panels (a,b), the Jaccard distances from
                population frequencies of all surviving haplotypes are
                shown, whereas panels (c,d) and (e,f) are based on the
                subsets $S_1= \{\mathrm{G238S}, \mathrm{R241P}, \mathrm{R164S},
                \mathrm{R164H}, \mathrm{E240G}, \mathrm{R241H}\} $ and $ S_2=
                \{\mathrm{G238S}, \mathrm{R241P}, \mathrm{R164S}\} $ of driver mutations, respectively. Error bars show the standard error of the mean for each data point.
	}
\label{fig:JaccardDistance}
\end{figure*}


In \Fref{fig:JaccardDistance}, we present the average Jaccard
distance measures for the three sets of mutations $S_0$, $S_1$ and $S_2$. 
The within-group (between-group) Jaccard distances are presented in left (right) panels.
The data for the largest antibiotic concentration are not shown as
there are simply no data except for the largest library size, and no trends can be inferred from a single data point.
Because each point is estimated from an average over three (pairs of)
EP-PCR pools, the error bars are seen to be quite substantial.
Nevertheless, several observations can be made from the figures.
First, the most noticeable difference between the two types of Jaccard
distance measures is the difference in vertical scale: the range of distances in the left panels spans from 0 to 0.4 whereas it runs from 0 to 1 in the right panels.
Between-group distances are generally considerably
larger than within-group distances, which shows the strong (possibly
dominant) role of the randomness induced by the EP-PCR 
in creating the diversity of selective outcomes.   
Using permutational multivariate analysis
of variance \cite{Anderson2001}, we checked that this difference is
indeed significant($P<0.001$).

Second, a pronounced difference in Jaccard distance measures is
observed for small library sizes ($ \le 10^3 $)  between the full set
of alleles $S_0$ (upper panels) and the sets of driver mutations $S_1$
and $ S_2 $ (middle and lower panels).
In particular, whereas the analysis of the full set of alleles
shows an apparent non-monotonic dependence of diversity on library size for
the two highest antibiotic concentrations, such a trend is absent when
only the driver mutations are considered. 
This behavior suggests that the large diversities of the small libraries are due to non-driver mutations. 
Since a small library size directly translates into a small number of
distinct alleles, the probability that multiple alleles carrying
driver mutations are present in one population is negligible for
sufficiently small libraries. Accordingly, two possible scenarios are
likely to occur. i) A single allele carrying beneficial mutations
exists initially in the population and eventually fixes. 
In this case, a small diversity among populations is expected as they share the same end-point haplotype.
ii) The population contains no allele carrying a driver mutation, in which case selection could very well result in extinction. However, this is clearly not always the case: at lower antiboitic concentrations\textemdash but in at least one population from a high antibiotic concentration (\Fref{fig:Library3})\textemdash  monomorphic populations carrying only the TEM-1 allele can sometimes show bacterial growth. Therefore, we attribute much of the high diversity at small library sizes to
the absence of strong selection effects and the presence of hitchhiking mutants of low or no resistance. Eliminating the contribution of these non-resistant mutants, we recover small values of the Jaccard distance measures for small
libraries and observe a robust general trend of increasing diversity with increasing library size.


Third, the behavior of the Jaccard distance measures depends only
weakly on the choices of driver mutations, i.e., the subsets $S_1$ and $S_2$ behave similarly. 
Thus the additional three mutations $\{\mathrm{R164H}, \mathrm{E240G}, \mathrm{R241H}\} $ contained in
$S_1$ but not in $S_2$ contribute only marginally to the trends of
the Jaccard distance measures. This suggests a simple picture in which
the fate of a population is determined only by the presence of the
driver mutations of largest effect, namely by G238S and subsequently
by R241P or R164S. Once such a driver mutation exists in the initial population, it will always constitute the dominating allele after selection.
To test this hypothesis, we introduce a mathematical model that relates the
probability of sampling haplotypes carrying driver mutations in the
initial population to the allelic composition of the final population.
The model contains a single parameter $\lambda$ for the mean number of
mutations in the TEM allele carried by a randomly chosen cell in the initial population. Different variants of the model can be considered depending on the choice of driver mutations. Here, we discuss the two cases $ S_2= \{\mathrm{G238S},
\mathrm{R241P}, \mathrm{R164S}\} $ and $ S_3= \{\mathrm{G238S}\}$
which display a clear contrast in the statistics. Further details of the model are provided in the Methods section.

If the simple selection scenario described above is correct, the probability that the initial population contains a certain set of driver mutations should explain the observed presence-absence patterns of those mutations in the final population. 
Therefore, an estimate of $\lambda$ can be made by maximizing the likelihood function from the observed data.
Finally, by comparing this estimate with the true value $\lambda =
1.5$, we may evaluate the validity of the scenario.
For the two sets of driver mutations $S_2$ and $S_3$, we found the estimates within $95\%$ confidence level to be $\lambda = 1.07 \pm 0.32$ and $\lambda = 1.81 \pm 0.82$, respectively.
These estimates of $\lambda$ are close to the true value of $1.5$, which confirms our hypothesis that the dominating factor in the selection experiment is the persistence of strong driver mutations and that the experiments are actually predictable given knowledge about the presence of driver mutations.
The smaller confidence interval obtained for $S_2$ is expected
because the number of final populations on which the analysis is based
is three times larger compared to $S_3$, which makes it puzzling that the estimate fails to explain the true value at the $95\%$ confidence level. 
We attribute this failure to possible cases in which multiple driver mutations are present and compete, which is neglected in the model.
Since the haplotypes R241P or R164S are suboptimal compared to G238S,
they may go extinct during the selection experiments in the presence
of G238S, which leads to an estimate of $\lambda$ that is too small. Furthermore, beneficiary TEM alleles\textemdash those with low resistance but high fitness in the absence of cefotaxime\textemdash might also play an important role late in selection. In this scenario, resistant alleles degrade cefotaxime, but the beneficiary allele reaches a high frequency in the final population due to their high fitness in the absence of cefotaxime. The occurrence of these social interactions could mask driver mutations, and therefore also result in too small estimates of $\lambda$. The ancestral TEM-1 allele could of course also be a beneficiary, if there are tradeoffs between resistance and fitness, perhaps explaining its repeated detection in the selected populations here. As a final remark, we emphasize that the analysis
has been performed under the assumption that epistasis is weak, which
is consistent with the analysis of the extinction patterns in Section \ref{sec:ExtinctionRates}. 
Although this seems to be largely true, there are several exceptions
such as haplotypes T114K+R164S+I173M+E240A and R164H+A172D 
that outcompete the haplotypes containing G238S even at the highest concentration. 


\section{Discussion and Conclusion}
We have considered the repeatability of selection, using mutant
libraries of different sizes and selection under different antibiotic
concentrations. We made a number of observations that are compatible
with previous results, and at the same time suggest that overall
patterns in the outcome of selection can be predicted in this
system. First, we observed higher rates of extinction in small
libraries and at higher antibiotic concentrations. Fitting of a simple
model to the data suggests that these extinction patterns can be
explained by the sampling of beneficial mutations and hard selection. Second, most alleles found in surviving populations contain known large-effect
mutations. When we modeled mutation and selection, a model with only
the three known largest-effect mutations - G238S, R241P and R164S - was best supported by the data. In this analysis, the effects of other mutations and epistasis are completely ignored, again suggesting that the outcome of short selection experiments is also predictable at the genotypic level. The evolution of TEM-1 is expected to be predictable, based on the heavy-tailed DBFE in which large-effect mutations are over-represented \cite{Schenk2012}, and other studies have highlighted the repeatability of evolution outcomes over multiple rounds of mutation and selection \cite{Barlow2002,Salverda2011}. Our results suggest that knowledge about mutational bias and the frequency at which large-effect driver mutations will occur are important for general predictions on the outcome of selection.

On the other hand, our work also highlights several complications with understanding the repeatability of evolution. First, in one case physical replicates fixed different mutations which would be expected to initiate two distinct evolutionary trajectories (G238S and R164S), highlighting the importance of stochastic effects, probably occurring early during selection, for the predictability of evolution. Since models are unlikely to capture such occurrences, these idiosyncratic outcomes will probably contribute to uncertainty in evolutionary predictions. Second, the increased Jaccard distances for large libraries was also unexpected, and suggests that predicting the specific driver mutations that will be present in selected populations is complex.

A non-monotonic relationship between population size and the
repeatability of evolution has been predicted in theoretical work
\cite{Szendro2013,Ochs2015}, including decreases in repeatability in
large populations due to the sampling of rare but highly beneficial
combinations of multiple mutations. Here we could not show the existence of such a non-monotonic relationship due to the large variation between biological replicates when the full dataset is considered (\Fref{fig:JaccardDistance}). Nevertheless, the data do suggest there may be such a trend, especially at cefotaxime concentrations $\geq 128$ ng/mL cefotaxime. Moreover, when only driver mutations are considered, there is a decrease in repeatability for large libraries, reminiscent of model predictions of lower repeatability in very large populations \cite{Szendro2013,Ochs2015}. One might reasonably expect that if smaller library sizes were taken and only driver mutations would be considered, repeatability would also decrease. Therefore, although we have not formally shown the existence of a non-monotonic trend here, both our complete and reduced datasets do suggest that such a relationship might exist between TEM library size and the repeatability of evolution.

What mechanisms underlie the decrease in the repeatability of evolution for large library sizes, suggested by our work? Recall that this trend has been observed for both within- and between-pool repeatability (\Fref{fig:JaccardDistance}). The low within-pool repeatability is probably not due to the sampling of different mutations in each replicate population. These replicates are all performed with alleles derived from the same pool and library. The starting population in these replicates represents a large sample ($\approx 10^6$ cells), in which each variant is, on average, present in a large number of cells that varies between $10^4$ in the smallest libraries to 10 in the largest libraries. This means that the vast majority of variants will be sampled for each replicate. However, for the largest libraries the number of cells carrying a given allele will be rather small and alleles might therefore go extinct regardless of their resistance, as the probability of survival is still small for resistant cells, suggesting a stochastic loss of beneficial alleles early in the selection process might account for low repeatability. A simple quantitative analysis presented in Section \ref{sec:SurvivalProbability} refutes this hypothesis and shows that the total survival probability always increases with library size. Whilst the mechanism causing low within-pool repeatability therefore remains elusive, our results suggest that it is events in the selection process that cause the low repeatability in large populations, and not genetic drift in the form of random sampling or extinction of beneficial alleles at the beginning of selection.

On the other hand, between-pool repeatability is lower than within-pool repeatability for large library sizes (\Fref{fig:JaccardDistance}), suggesting that mutational supply also plays a role in reducing repeatability. Indeed, in one library a high resistance TEM allele dominated all selected populations at all antibiotic concentrations (i.e., the G238S L40* allele in the $10^5$ size of library B), whilst in the other libraries such "all-round winners" are not present. Moreover, we also showed that alleles carrying multiple mutations can confer higher resistance than alleles with only a large-effect mutation (\Tref{tab:Resistance}). Highly beneficial alleles with multiple mutations clearly exist, but are probably rare and only occasionally sampled even in the largest libraries. There are therefore clear parallels between our work and model predictions on the decrease repeatability at large population sizes \cite{Szendro2013,Ochs2015}. However, as we performed a single round of selection on standing genetic variation, the putative scarcity of these alleles can in itself account for the observed patterns and the accessibility of evolutionary trajectories will not play a role.

\section{Methods}
\subsection{Media}
LB medium is 10 g/L Bacto tryptone, 5 g/L yeast extract, and 10 g/L NaCl. LB-agar is LB medium with 15 g/L agarose added before autoclaving. Mueller-Hinton (MH; Becton-Dickenson) medium was prepared according to the manufacturer’s instructions. SOC medium was prepared with 20 g/L Bacto Tryptone, 5 g/L yeast extract, 2.5 mM KCl, 10 mM NaCL, 10 mM MgCl$_2$, 10 mM MgSO$_4$ and 20 mM Glucose.

\subsection{Bacterial strains and plasmids}
\textit{E. coli} strain DH5αE (Invitrogen) was used for all in vivo steps of the experiment. Plasmid pACSE3 \cite{Barlow2002} was used as a vector for all cloning and transformation steps, and is referred to as pACTEM when it contains a TEM allele. The vector contains a tetracycline resistance gene, and tetracycline was added to all LB media (15 ug/mL) to retain only plasmid containing cells. TEM expression was induced by adding 50 μM isopropyl-b-D-thiogalactopyranoside (IPTG) to the medium.

\subsection{Mutant library construction \&  library size estimation}
Three TEM mutant libraries were constructed with an error-prone polymerase chain reaction (PCR), using the GeneMorph II Random Mutagenesis Kit (Agilent Technologies). We aimed to obtain a low mutation rate (0-4 mutations/Kb) by using 100 ng of template DNA (which equates 565 ng pACTEM), and running the PCR reaction for 25 cycles in a 50 uL reaction volume \cite{Salverda2011}. Primers used were P3: 5’-TCATCCGGCTCGTATAATGTGTGGA and P4: 5’-ACTCTCTTCCGGGCGCTATCAT \cite{Barlow2002}. The PCR-product was purified with a NucleoSpin Gel and PCR Clean-up kit (Macherey-Nagel), and digested with restriction endonucleases BspHI, SacI, and DpnI (New England Biolabs). 

TEM alleles were ligated into the pACSE3 vector overnight at 4 ºC with 3:1 insert:vector ratio, using T4 Ligase (Promega). Two uL of the ligate was used for electroporation, and the transformed cells were suspended in SOC-medium and incubated at 37ºC for 90 min to recover. A dilution of the electroporated cell cultures was plated out on LB agar supplemented with tetracycline (15 mg/L) and incubated overnight at 37ºC. The remainder of the recovered transformants were stored at 4 ºC to prevent further growth, with mild agitation. The number of transformants was estimated from counting colonies on the plates. The fraction of transformants containing a plasmid with TEM inserted was determined from 12 P3/P4 colony PCR reactions to verify the presence of the full-length TEM gene. Effective library size was the number of transformants multiplied by the fraction of transformants containing TEM.  

\subsection{Library enrichment, antibiotic selection and sequencing}
From the effective library sizes, volumes containing approximately $10^2, 10^3, 10^4$ and $10^5$ cells with TEM were calculated. These volumes were then used to inoculate 100 mL bottles with 50 mL LB medium supplemented with 15 μg/ml tetracycline, and incubated overnight at 37 ºC without agitation to complete library enrichment. Note since the library size is determined by sampling prior to library amplification, smaller libraries are not nested within larger libraries. On the other hand, since mutations can occur at any cycle during the EP-PCR, the occurrence of mutations within a single PC reaction is likely to be coordinated (i.e., samples drawn from the same EP-PCR will be more alike than samples drawn from independent reactions). Selection was performed in nine mL of Muller-Hinton broth containing 64, 128, 256, or 512 ng/mL cefotaxime (Duchefa), and inoculated with $10^6$ cells, as determined by absorbance (OD$_{600}$) measurements on the amplified libraries. These cultures were incubated at 37 ºC without agitation for 48 h. The cultures were then briefly spun down, and whether a culture had grown was determined by visually inspecting tubes for a pellet. Cells were then suspended in 450 μL LB containing 15\% glycerol, and stored at -80ºC. A 100$\times$ dilution of these suspensions was used as a template for P3/P4 colony-PCR in a 30 μL reaction volume. The PCR products were purified and Sanger sequenced, using the P4 primer (Eurofins Genomics, Germany). To characterize our libraries, we also plated cells from the three $10^5$ size amplified libraries on LB with 15 μg/ml tetracycline, and then amplified and sequenced TEM for 6 individual colonies.

\subsection{MIC and competitions}
Populations of interest were plated on LB agar with 15 ug/ml tetracycline, and we then isolated plasmid DNA from overnight cultures of clones. Plasmid DNA was transformed into new DH5αE cells. MIC assays were performed in 250 μL Muller Hinton in 96-well plates, with 50 μM IPTG and 0.0625 to 64 μg/mL cefotaxime, with 2-fold dilutions. Initial cell density was the same as for the selection on libraries ($10^5$ cells per mL). The plates were incubated at 37 ºC for 48 h, and the optical density at 600 nm (OD$_{600}$) was measured using a Victor$^3$ plate reader (Perkin-Elmer). OD$_{600}$ > 0.05 were considered positive to estimate the MIC.

Direct competitions were performed in 5 mL Muller Hinton, with 50 μM IPTG and either no antibiotics or 0.256 μg/mL cefotaxime. Exponential phase cells were mixed to give a 1:1 ratio of competitors, at an initial density of $10^3$ or $10^5$ cells per mL, and cultures were incubated at 37 ºC for 48 h. PCR with the P3 and P4 primers was performed directly on 0.5 μL of the inoculum mixture and the final populations, and the PCR products were Sanger sequenced. The frequency of the different TEM alleles was estimated based on the peak height at polymorphic sites (see also Analysis of Sanger sequences). For competitions between alleles differing at more than one nucleotide position, the mean frequency based on all polymorphic positions was used. We estimated the number of cells at the start and end of competitions from OD$_{600}$ measurements and the frequency of the two variants, and then took the ratio of their Malthusian parameters as a measure of relative fitness, \textit{W}.

\subsection{Analysis of Sanger sequences}
DNA sequence files were imported in CodonCode Aligner (v 6.0.2), trimmed, and aligned to the TEM-1 reference sequence. Initially, mutations were visually scored in the chromatograms, and double peaks were assumed to represent polymorphisms in the sampled populations. Next, we measured the height of peaks for all polymorphic sites using ImageJ (National Institutes of Health), and estimated the frequency of different bases from relative peak heights. We disregarded any bases with a frequency below 0.1, including those corresponding to the ancestral TEM-1. For those populations with more than one polymorphic site, we inferred heuristically the alleles present (i.e., haplotypes consisting of none, one or multiple single-nucleotide mutations) by considering the variation between replicates of the same dilution of the library. Pearson correlations and principle component analysis on the mutation frequencies were used to further support these intuitions. For polymorphic populations, we also sequenced between 4 and 10 clones from one physical replicate to test whether our haplotype inferences were correct. TEM-1 was also included, although we could not distinguish it from a possible consortium of many different variants at low frequencies.

\subsection{Extinction model}
For every TEM allele present in a library, we assume that the total number of single-nucleotide mutations per allele introduced by the EP-PCR is Poisson distributed, with a mean of $ \Lambda $ and a realization of this distribution being $ \lambda $. Based on sequencing of the TEM alleles from clones from the enriched library (i.e., prior to selection), we can estimate that $ \Lambda = 1.5 $ and showed that a Poisson distribution is a good approximation of the empirically observed distribution (See \Fref{fig:EPLibraries}). Next, we assume that all mutations are either beneficial or non-beneficial, and that the distribution of beneficial mutations per TEM allele is binomially distributed, with the probability that a mutation is beneficial being $ \beta $ and the total number trials being the mutations on a given allele, $ \lambda $. From a previous study \cite{Schenk2012},  we estimated that $ \beta  = 87/(3*861) \sim 0.034 $, given a minimum estimate of 87 mutations that increase the resistance of TEM-1 to cefotaxime, and the length of the TEM open reading frame. We assume that all non-beneficial mutations are deleterious, having a resistance equal to cells carrying the empty vector pACSE3 (IC$ _{99.99} $ = 33 ng/ml), which has a lower resistance than TEM-1 (pACTEM IC$ _{99.99} $ = 52) \cite{Schenk2012}[35]. For beneficial mutations, we assume that their resistance is a randomly drawn value – with replacement, as not all beneficial mutations are known – from the empirical heavy-tailed distribution of beneficial fitness effects of 48 single-nucleotide mutations found for TEM-1 \cite{Schenk2012}. The sampling of values from this empirical distribution is weighted to reflect mutational bias, given the mutational spectrum of the GeneMorph II kit is known \cite{Anonymous2015} (see also \Fref{fig:ExitinctionModel}c).

Next, we need to make assumptions about epistasis, since there can be multiple single-nucleotide mutations in a single TEM allele. For Models 1 and 2, we assume that the resistance effects of multiple mutations are additive, following the definition given by Sanjuan and Elena \cite{Sanjuan2006}. The change in log of resistance for a mutated TEM allele ($ \Delta \ln r_a $) is the sum of the differences in resistance compared to TEM-1 for the log-transformed IC$ _{99.99} $ values ($ \Delta \ln r_i $). Next, to allow for epistasis in Models 3 and 4, we assume there is a fixed magnitude of epistatic interactions for all mutations introduced into TEM, with a value $ \epsilon $. Finally, in Models 2 and 4 we consider the possibility that it might not be possible to use our resistance measurements for single-nucleotide mutations (IC$ _{99.99} $) to infer the exact maximum antibiotic concentration at	 which a mutant can grow by itself in the selection experiments described here. Such an effect is likely, since the IC$ _{99.99} $ measurements are made on plates and not in liquid culture, and the initial density of cells is different. We therefore include the possibility to rescale resistance by including a constant $ \rho $ in the estimation of the resistance of a TEM allele. Hence for alleles carrying one or more mutations: 
\begin{equation}
\Delta \ln r_a = \rho +_ \epsilon (\Lambda-1) + \sum_{i=1}^{\Lambda} \Delta \ln r_i
\end{equation}
For alleles with no mutations $ \Delta \ln r_a = \rho $.
For a library size $ L $, we draw $ L $ TEM variants and determine $r_a$ for all the drawn alleles. We assume there is hard selection and no ‘social interactions’ (e.g., less resistant variants growing after more resistant variants have degraded the antibiotic in the environment). If $r_a$ is greater than or equal to the concentration of the antibiotic cefotaxime present for any of these $ L $ variants present, that population will not go extinct. By iterating this process a large number of times (i.e., simulating selection on many different EP-PCR libraries), we can generate a prediction for the frequency of extinction for all the conditions used in the experiment. For small library sizes ($ L < 10^5 $) 1000 iterations were performed, whereas for large populations ($ L \ge 10^5 $) 250 iterations were performed.

To fit this models to the data, we calculated the negative log likelihood (NLL) based on the binomial likelihoods obtained by comparing the number of observed extinctions to the model prediction for each combination of $L$ and cefotaxime concentration. However, the model can predict that all lineages survive or go extinct, and if the data are not identical to the model prediction (only survival or extinction observed, respectively), a likelihood cannot be calculated. To always obtain an indication of model fit and make model fitting tractable, we therefore used the Laplace binomial point estimator (LPBE: $ [x+1]/[n+2] $) for the predicted model frequency of extinction. Note that the number of iterations performed for estimating the frequency of extinction will have an effect on NLL values, since we are using the LBPE instead of $ x/n $. For Models 2 and 3 there is one free parameter, and there are two for Model 4: the constant to scale resistance $ \rho $ and the magnitude of epistatic interactions $ \epsilon $. To estimate these parameters, grid searches were performed to minimize the NLL. Searches were initially performed over a broad range of parameter estimates: (-4, 4) for $ \rho $ and $ \epsilon $ with an interval of 0.5. Subsequently grid searches were performed over three smaller spaces where NLL values tended to be low: $ \rho = (-3, -0.5), \epsilon = (-0.5, 1); \rho = (-1.7, -1.1), \epsilon = (-1.5, -0.6); \rho = (-1, -0), \epsilon = (-4, -0.6) $; all with interval 0.1. Finally, because the model is stochastic and the number of simulation runs during the grid search was not very large (computational constraint)s, for Models 1-4 the model was run 10 times with estimated parameters, and we used the median NLL value for model selection with AIC.

\subsection{Analysis of repeatability of selection}
In this subsection, we construct a model aimed at testing our hypothesis 
	that the fate of populations is \textit{a priori} determined by the presence of driver mutations. 
To this end, we first calculate the probability that the initial population contains haplotypes carrying driver mutations.
Note that this probability is dependent on the library size
and on the mean number $\lambda$ of single-nucleotide mutations of average haplotypes. 
Even though $\lambda$ is known to be $1.5$, it will be treated as an
unknown parameter and the maximum likelihood estimates of $\lambda$ will be
compared to the true value to assess the validity of the hypothesis.

To proceed, we consider the simpler subproblem of determining $P_k(x,
y)$, the probability that a haplotype with $k$ substitutions contains
a specific nucleotide substitution from $x$ to $y$ at a specific
position of the TEM-1 plasmid, where $x, y \in
\mathcal{N} = \{\mathrm{a, c, g, t}\}$ .
In this minimal model, we assume that the probability of a
substitution 
does not depend on its spatial position
but only on the types of nucleotides involved due to the known mutational biases. According to \cite{Anonymous2015}, the normalized relative weight $\eta(x, y) $ of a mutation from $x$ to $y$ is given by the matrix
\begin{eqnarray}
\begin{array}{ccc}
\eta(\Nuc{a}, \Nuc{c}) = 0.0249   &  \eta(\Nuc{a}, \Nuc{g}) = 0.0927  &  \eta(\Nuc{a}, \Nuc{t}) = 0.0151  \\ 
\eta(\Nuc{c}, \Nuc{a}) = 0.0747   &  \eta(\Nuc{c}, \Nuc{g}) = 0.0217  &  \eta(\Nuc{c}, \Nuc{t}) = 0.1351 \\ 
\eta(\Nuc{g}, \Nuc{a}) = 0.1351   &  \eta(\Nuc{g}, \Nuc{c}) = 0.0217  &  \eta(\Nuc{g}, \Nuc{t}) = 0.0747 \\ 
\eta(\Nuc{t}, \Nuc{a}) = 0.0151   &  \eta(\Nuc{t}, \Nuc{c}) = 0.0927  &  \eta(\Nuc{t}, \Nuc{g}) = 0.0249,
\end{array} 
\end{eqnarray}
where it is easily checked that $\sum_{x , y \in \mathcal{N}} \eta(x,y) = 1$. 
Additionally, we define $\eta(x)$ to be the sum of weights of
mutations of $x$ to any other nucleotide, i.e., $ \eta(x) = \sum_{y \in \mathcal{N}}  \eta(x,y)$.
Given these quantities, one may write
\begin{eqnarray}
P_k(x, y) &= \frac{ \eta(x,y) \times W_{k-1}(l_\Nuc{a} - \delta_{x, \Nuc{c}}, l_\Nuc{c}- \delta_{x, \Nuc{a}}, l_\Nuc{g}- \delta_{x, \Nuc{g}}, l_\Nuc{t}- \delta_{x, \Nuc{t}})}{W_k(l_\Nuc{a}, l_\Nuc{c}, l_\Nuc{g}, l_t)},
\end{eqnarray}
where $\delta_{i,j}$ is the Kronecker delta,  $l_x$ is the number of sites of type $x$ in the original TEM-1 gene ($l_\Nuc{a} = 223, l_\Nuc{c} = 203, l_\Nuc{g} = 222, l_\Nuc{t} = 213$) and
\begin{eqnarray}
W_k(l_1, l_2, l_3, l_4) &= \parbox[t]{0.55\linewidth}{sum of weights of choosing $k$ substitutions out of $ l_1, l_2, l_3$ and  $ l_4 $ nucleotides}\nonumber \\
&= \sum_{\scriptstyle n_1,n_2,n_3, n_4 \ge 0 \atop\scriptstyle n_1 + n_2 + n_3 + n_4 = k} \prod_{x \in \mathcal{N}} \binom{l_x}{n_x} \eta(x)^{n_x}.
\end{eqnarray}
As discussed in the previous Section \ref{sec:StatisticsLibs} and elsewhere \cite{Sun1995},
the total number of single-nucleotide mutations is known to follow a
Poisson distribution with mean $\lambda$, so that the probability of finding a cell with $k$ substitutions is $q_k = \lambda^k e^{-\lambda} / k! $. 
Hence, the expected number of haplotypes with $k$ substitutions in the initial population of library size $L$ is $ L q_k $ with a standard deviation of order $\sqrt{L}$. 
Since we are considering fairly large library sizes, the fluctuation
of this number around its mean is ignored in the rest of the analysis.

Now, we are ready to construct the likelihood function. 
For each error-prone PCR pool and for each library size, the presence
and absence of driver mutations can be easily determined from the
haplotype tables in \Fref{fig:Library1}-\ref{fig:Library3}. If a specific driver mutation realized by the substitution $x$ to $y$ is not observed in the corresponding haplotype table, the probability 
\begin{eqnarray}
\parbox[t]{0.32\linewidth}{Prob(Not found from $x$ to $y$)} = \prod_{k=0}^{\infty} \left(
	1 - P_k(x,y)
\right)^{L q_k},
\end{eqnarray}
is assigned to this population.
Note that this probability implicitly depends on $\lambda$ via the $ q_k $.
Similarly, if a haplotype of $k$ substitutions carries this specific
driver mutation, we assign it the probability
\begin{eqnarray}
\parbox[t]{0.35\linewidth}{Prob(found at $k$ from $x$ to $y$)} = 1 - \left(
1 - P_k(x,y)
\right)^{L q_k}.
\end{eqnarray}
Additionally, once such haplotypes are found, no additional constraints for different numbers of substitutions are imposed. 
This is based on the fact that our hypothesis only assumes strong
selection against cells without driver mutations, and therefore no
additional information is gained about the competition between two haplotypes with the same driver mutation.
Finally, after assigning probabilities for all combinations of EP-PCR pools and library sizes, the likelihood function is constructed by taking the product of all probabilities.
Then, the maximization of the likelihood function yields an estimate
of the optimal $\lambda$ based on the presence-absence data, and
the confidence intervals are calculated numerically from the local curvature of the log-likelihood function around the optimal value.

\begin{figure*}[t]
	\centering
	\includegraphics[width=0.8\textwidth]{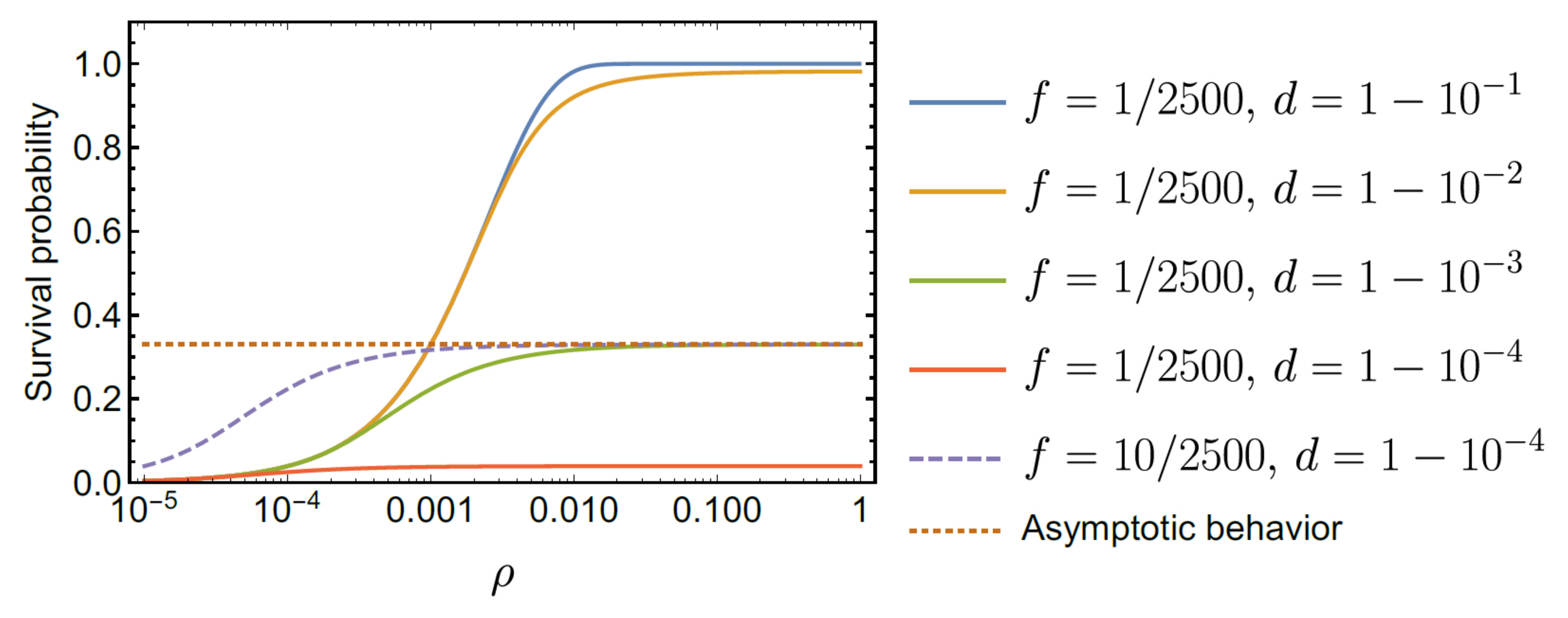}
	\caption{
		Survival probabilities obtained from \Eref{SurvivalProbability} for various choices of the sampling probability $f$ and the single-cell extinction
		probability $d$, with the initial number of cells in the competition being $N=10^6$.
		For the first four curves, the sampling probability is
		set to be $1/2500$, which is close to the probability
		of randomly sampling one of the $ 2583 = 861 \times 3
		$ single nucleotide substitutions of TEM-1. The
		library sizes used in the experiment correspond to
		$\rho=10^{-4}, 10^{-3}, 10^{-2}$ and $10^{-1}$.
		As the single-cell extinction probability $d$ increases, the survival probability is shown to decrease for sufficiently large $\rho$ whereas the small $\rho$ behavior (for small library sizes) is independent of $d$. 
		The dashed line shows the effect of increasing the
		initial frequency $f$ of the allele. When $d$ is close
		to 1, the asymptotic plateau value of the survival
		probability indicated by the dotted line depends only on the product $f \times (1-d)$ (\Eref{SurvivalProbability2}).
	}
	\label{fig:SurvivalProbability}
\end{figure*}

\subsection{Allelic survival probability increases monotonically with library size}
\label{sec:SurvivalProbability}
In our experimental setting, there are two competing factors that
determine the probability that an allele is observed after selection,
termed \textit{survival probability} in the following: i) 
haplotypes carrying the allele should be initially present in the
library before selection and ii) they should survive the selection step. 
Obviously, because of factor i), one expects that the survival
probability increases with library size $L$, because sampling more cells gives a larger chance to find a certain allele.
However, the subsequent amplification step that normalizes the
different library sizes to the same population size $N=10^6$ at the
beginning of the selection experiment introduces an opposing
effect. Because the number of cells is multiplied by a factor $1/\rho := N/L$, the number of cells carrying a given allele that was sampled
is smaller in the larger libraries, which implies an enhanced chance
of extinction. To understand how these competing effects play out, here we present a simple calculation which shows that the survival probability always increases monotonically with library size.

Within a given EP-PCR pool, each allele is characterized by two
parameters: the sampling probability $f$ and the single-cell extinction
probability during the selection process $d$. 
Under the uniform sampling assumption, the probability of sampling $n$ cells having this allele is binomially distributed with parameter $f$.
The single cell survival probability is the probability that not all of the
$n/\rho$ cells that are present in the beginning of the selection step
go extinct. Summing over $n$, the probability of observing the allele
of interest after selection is thus given by
\begin{equation}
S(\rho) = \sum _{n=1}^{\infty } \binom{L}{n} f^n (1-f)^{L-n}
(1-d^{\frac{n}{\rho}}) = 1 - (1-f + f d^{1/\rho})^{N \rho}.
\label{SurvivalProbability}
\end{equation}
The behavior of $S(\rho)$ for various parameters that resemble the
experimental conditions is sketched in
\Fref{fig:SurvivalProbability}.
The survival probability increases monotonically and
reaches a plateau value for large $\rho$ which is approximately given
by 
\begin{equation}
\label{SurvivalProbability2}
S(\rho) \approx 1 - e^{N f(1-d)}
\end{equation}
provided $d$ is close to 1. 


To prove that $S(\rho)$ is indeed monotonically increasing over the
entire range of $\rho$, $0 \le \rho \le 1$, it is sufficient to show that $\rho \ln (1-f+fd^{1/\rho})$ is a decreasing function of $\rho$. 
After taking the derivative with regard to $\rho$, this condition is recast as
\begin{eqnarray}
\label{Inequality}
	\ln \left(1-f+f d^{\frac{1}{\rho }}\right)-\frac{f d^{\frac{1}{\rho }} \ln (d^{1/\rho})}{\left(1-f+ f d^{\frac{1}{\rho }}\right)} \stackrel{!}{\le} 0,
\end{eqnarray}
for any choice of $\rho$, $d$ and $f$.
By introducing a variable $q = f d^{1/\rho}$ and multiplying both
sides by $(1-f +q)$, we find the condition
\begin{eqnarray}
	G(q) := (1-f+q)\ln (1 - f + q)  - q \ln q + q \ln f \stackrel{!}{\le} 0,
\end{eqnarray}
where $q$ now takes values in the range $0 \leq q \leq f$. 
It is easy to check that $G'(q) > 0$ for $0 < q < f$.
Together with the fact that $G(f) = 0$ at the right boundary, the
inequality is established.

\ack

We thank Diego Pesce and Bertha Koopmanschap for valuable advice.


\section*{References}

\bibliographystyle{unsrt}
\bibliography{me}

\begin{thebibliography}{10}

\bibitem{Visser2014}
J.~A. G.~M. de~Visser and Joachim Krug.
\newblock Empirical fitness landscapes and the predictability of evolution.
\newblock {\em Nature Reviews Genetics}, 15:480--490, 2014.

\bibitem{Dyson1920}
F.~W. Dyson, A.~S. Eddington, and C.~Davidson.
\newblock A determination of the deflection of light by the
  sun{\textquoteright}s gravitational field, from observations made at the
  total eclipse of may 29, 1919.
\newblock {\em Philosophical Transactions of the Royal Society of London A:
  Mathematical, Physical and Engineering Sciences}, 220(571-581):291--333,
  1920.

\bibitem{Buckling2009}
Angus Buckling, R.~Craig~Maclean, Michael~A. Brockhurst, and Nick Colegrave.
\newblock The beagle in a bottle.
\newblock {\em Nature}, 457(7231):824--829, February 2009.

\bibitem{Daeschler2006}
Edward~B. Daeschler, Neil~H. Shubin, and Farish~A. Jenkins.
\newblock A devonian tetrapod-like fish and the evolution of the tetrapod body
  plan.
\newblock {\em Nature}, 440(7085):757--763, April 2006.

\bibitem{Soria-Carrasco2014}
V{\'\i}ctor Soria-Carrasco, Zachariah Gompert, Aaron~A. Comeault, Timothy~E.
  Farkas, Thomas~L. Parchman, J.~Spencer Johnston, C.~Alex Buerkle, Jeffrey~L.
  Feder, Jens Bast, Tanja Schwander, Scott~P. Egan, Bernard~J. Crespi, and
  Patrik Nosil.
\newblock Stick insect genomes reveal natural selection{\textquoteright}s role
  in parallel speciation.
\newblock {\em Science}, 344(6185):738--742, 2014.

\bibitem{Yeaman2016}
Sam Yeaman, Kathryn~A. Hodgins, Katie~E. Lotterhos, Haktan Suren, Simon Nadeau,
  Jon~C. Degner, Kristin~A. Nurkowski, Pia Smets, Tongli Wang, Laura~K. Gray,
  Katharina~J. Liepe, Andreas Hamann, Jason~A. Holliday, Michael~C. Whitlock,
  Loren~H. Rieseberg, and Sally~N. Aitken.
\newblock Convergent local adaptation to climate in distantly related conifers.
\newblock {\em Science}, 353(6306):1431--1433, 2016.

\bibitem{Natarajan2016}
Chandrasekhar Natarajan, Federico~G. Hoffmann, Roy~E. Weber, Angela Fago,
  Christopher~C. Witt, and Jay~F. Storz.
\newblock Predictable convergence in hemoglobin function has unpredictable
  molecular underpinnings.
\newblock {\em Science}, 354(6310):336--339, 2016.

\bibitem{Luksza2014}
Marta Luksza and Michael L\"assig.
\newblock A predictive fitness model for influenza.
\newblock {\em Nature}, 507(7490):57--61, March 2014.

\bibitem{Barlow2002}
Miriam Barlow and Barry~G. Hall.
\newblock Predicting evolutionary potential: In vitro evolution accurately
  reproduces natural evolution of the tem beta-lactamase.
\newblock {\em Genetics}, 160(3):823--832, 2002.

\bibitem{MacLean2010}
R~Craig MacLean, Alex~R Hall, Gabriel~G Perron, and Angus Buckling.
\newblock The population genetics of antibiotic resistance: integrating
  molecular mechanisms and treatment contexts.
\newblock {\em Nature Reviews Genetics}, 11:405--414, June 2010.

\bibitem{Wright1932}
S.~Wright.
\newblock The roles of mutation, inbreeding, crossbreeding, and selection in
  evolution.
\newblock {\em Proc. 6th Int. Cong. Genet.}, 1:356--366, 1932.

\bibitem{Weinreich2006}
Daniel~M. Weinreich, Nigel~F. Delaney, Mark~A. DePristo, and Daniel~L. Hartl.
\newblock Darwinian evolution can follow only very few mutational paths to
  fitter proteins.
\newblock {\em Science}, 312:111--114, 2006.

\bibitem{Poelwijk2007}
F.~J. Poelwijk, D.~J. Kiviet, D.~M. Weinreich, and S.~J. Tans.
\newblock Empirical fitness landscapes reveal accessible evolutionary paths.
\newblock {\em Nature}, 445:383--386, 2007.

\bibitem{Poelwijk2011}
Frank~J. Poelwijk, Sorin Tănase-Nicola, Daniel~J. Kiviet, and Sander~J. Tans.
\newblock Reciprocal sign epistasis is a necessary condition for multi-peaked
  fitness landscapes.
\newblock {\em Journal of Theoretical Biology}, 272(1):141 -- 144, 2011.

\bibitem{Salverda2010}
Merijn~L.M. Salverda, J.~Arjan~G.M. De~Visser, and Miriam Barlow.
\newblock Natural evolution of tem-1 β-lactamase: experimental reconstruction
  and clinical relevance.
\newblock {\em FEMS Microbiology Reviews}, 34(6):1015, 2010.

\bibitem{Salverda2011}
Merijn L.~M. Salverda, Eynat Dellus, Florien~A. Gorter, Alfons J.~M. Debets,
  John van~der Oost, Rolf~F. Hoekstra, Dan~S. Tawfik, and J.~Arjan G.~M.
  de~Visser.
\newblock Initial mutations direct alternative pathways of protein evolution.
\newblock {\em PLOS Genetics}, 7(3):1--11, 03 2011.

\bibitem{Schenk2013}
Martijn~F. Schenk, Ivan~G. Szendro, Merijn~L.M. Salverda, Joachim Krug, and
  J.~Arjan~G.M. de~Visser.
\newblock Patterns of epistasis between beneficial mutations in an antibiotic
  resistance gene.
\newblock {\em Molecular Biology and Evolution}, 2013.

\bibitem{Dellus-Gur2015}
Eynat Dellus-Gur, Mikael Elias, Emilia Caselli, Fabio Prati, Merijn~L.M.
  Salverda, J.~Arjan~G.M. de~Visser, James~S. Fraser, and Dan~S. Tawfik.
\newblock Negative epistasis and evolvability in tem-1 β-lactamase—the thin
  line between an enzyme's conformational freedom and disorder.
\newblock {\em Journal of Molecular Biology}, 427(14):2396 -- 2409, 2015.

\bibitem{Blount2012}
Zachary~D. Blount, Jeffrey~E. Barrick, Carla~J. Davidson, and Richard~E.
  Lenski.
\newblock Genomic analysis of a key innovation in an experimental escherichia
  coli population.
\newblock {\em Nature}, 489(7417):513--518, September 2012.

\bibitem{Tenaillon2012}
Olivier Tenaillon, Alejandra Rodr{\'\i}guez-Verdugo, Rebecca~L. Gaut, Pamela
  McDonald, Albert~F. Bennett, Anthony~D. Long, and Brandon~S. Gaut.
\newblock The molecular diversity of adaptive convergence.
\newblock {\em Science}, 335(6067):457--461, 2012.

\bibitem{Kryazhimskiy2014}
Sergey Kryazhimskiy, Daniel~P. Rice, Elizabeth~R. Jerison, and Michael~M.
  Desai.
\newblock Global epistasis makes adaptation predictable despite sequence-level
  stochasticity.
\newblock {\em Science}, 344(6191):1519--1522, 2014.

\bibitem{Fisher1930}
R.~A. Fisher.
\newblock {\em The Genetical Theory of Natural Selection}.
\newblock Clarendon Press, Oxford, 1930.

\bibitem{Muller1932}
H.~J. Muller.
\newblock Some genetic aspects of sex.
\newblock {\em Am. Nat.}, 66:118--138, 1932.

\bibitem{Crow1965}
J.~F. Crow and M.~Kimura.
\newblock Evolution in sexual and asexual populations.
\newblock {\em Am. Nat.}, 99:439--450, 1965.

\bibitem{Gerrish1998}
P.~J. Gerrish and R.~E. Lenski.
\newblock The fate of competing beneficial mutations in an asexual population.
\newblock {\em Genetica}, 102-103:127--144, 1998.

\bibitem{Lachapelle2015}
Josianne Lachapelle, Joshua Reid, and Nick Colegrave.
\newblock Repeatability of adaptation in experimental populations of different
  sizes.
\newblock {\em Proceedings of the Royal Society of London B: Biological
  Sciences}, 282(1805), 2015.

\bibitem{Rozen2008}
Daniel~E. Rozen, Michelle G. J.~L. Habets, Andreas Handel, and J.~Arjan G.~M.
  de~Visser.
\newblock Heterogeneous adaptive trajectories of small populations on complex
  fitness landscapes.
\newblock {\em {PLoS} {ONE}}, 3(3):e1715, mar 2008.

\bibitem{Szendro2013}
Ivan~G. Szendro, Jasper Franke, J.~Arjan G.~M. de~Visser, and Joachim Krug.
\newblock Predictability of evolution depends nonmonotonically on population
  size.
\newblock {\em Proc. Nat. Acad. Sci. USA}, 110:571--576, 2013.

\bibitem{Ochs2015}
Ian~E. Ochs and Michael~M. Desai.
\newblock The competition between simple and complex evolutionary trajectories
  in asexual populations.
\newblock {\em BMC Evolutionary Biology}, 15(1):55, 2015.

\bibitem{Vogwill2016}
Tom Vogwill, Robyn~L. Phillips, Danna~R. Gifford, and R.~Craig MacLean.
\newblock Divergent evolution peaks under intermediate population bottlenecks
  during bacterial experimental evolution.
\newblock {\em Proceedings of the Royal Society of London B: Biological
  Sciences}, 283(1835), 2016.

\bibitem{Herron2013}
Matthew~D. Herron and Michael Doebeli.
\newblock Parallel evolutionary dynamics of adaptive diversification in
  escherichia coli.
\newblock {\em PLOS Biology}, 11(2):1--11, 02 2013.

\bibitem{Deris2013}
J.~Barrett Deris, Minsu Kim, Zhongge Zhang, Hiroyuki Okano, Rutger Hermsen,
  Alexander Groisman, and Terence Hwa.
\newblock The innate growth bistability and fitness landscapes of
  antibiotic-resistant bacteria.
\newblock {\em Science}, 342(6162), 2013.

\bibitem{Yurtsev2013}
Eugene~A Yurtsev, Hui~Xiao Chao, Manoshi~S Datta, Tatiana Artemova, and Jeff
  Gore.
\newblock Bacterial cheating drives the population dynamics of cooperative
  antibiotic resistance plasmids.
\newblock {\em Molecular Systems Biology}, 9(1), 2013.

\bibitem{Kirkwood1994}
T~B Kirkwood and C~R Bangham.
\newblock Cycles, chaos, and evolution in virus cultures: a model of defective
  interfering particles.
\newblock {\em Proceedings of the National Academy of Sciences of the United
  States of America}, 91(18):8685--8689, August 1994.

\bibitem{Zwart2013}
Mark~P Zwart, Gorben~P Pijlman, Josep Sardanyés, Jorge Duarte, Cristina
  Januário, and Santiago~F Elena.
\newblock Complex dynamics of defective interfering baculoviruses during serial
  passage in insect cells.
\newblock {\em Journal of Biological Physics}, 39(2):327--342, March 2013.

\bibitem{Artemova2015}
Tatiana Artemova, Ylaine Gerardin, Carmel Dudley, Nicole~M Vega, and Jeff Gore.
\newblock Isolated cell behavior drives the evolution of antibiotic resistance.
\newblock {\em Molecular Systems Biology}, 11(7):822--, July 2015.

\bibitem{Schenk2012}
Martijn~F. Schenk, Ivan~G. Szendro, Joachim Krug, and J.~Arjan G.~M. de~Visser.
\newblock Quantifying the adaptive potential of an antibiotic resistance
  enzyme.
\newblock {\em PLOS Genetics}, 8(6):1--11, 06 2012.

\bibitem{Sun1995}
Fengzhu Sun.
\newblock The polymerase chain reaction and branching processes.
\newblock {\em Journal of Computational Biology}, 2(1):63--86, January 1995.

\bibitem{Drummond2005}
D.~Allan Drummond, Brent~L. Iverson, George Georgiou, and Frances~H. Arnold.
\newblock Why high-error-rate random mutagenesis libraries are enriched in
  functional and improved proteins.
\newblock {\em Journal of Molecular Biology}, 350(4):806 -- 816, 2005.

\bibitem{Anonymous2015}
Anonymous.
\newblock {\em GeneMorph II random mutagenesis kit instruction manual}, 2015.

\bibitem{Sideraki2000}
Vera Sideraki, Wanzhi Huang, Timothy Palzkill, and Hiram~F Gilbert.
\newblock A secondary drug resistance mutation of tem-1 β-lactamase that
  suppresses misfolding and aggregation.
\newblock {\em Proceedings of the National Academy of Sciences of the United
  States of America}, 98(1):283--288, September 2000.

\bibitem{Firnberg2014}
Elad Firnberg, Jason~W. Labonte, Jeffrey~J. Gray, and Marc Ostermeier.
\newblock A comprehensive, high-resolution map of a gene’s fitness landscape.
\newblock {\em Molecular Biology and Evolution}, 31(6):1581--1592, 2014.

\bibitem{Whittaker1960}
R.~H. Whittaker.
\newblock Vegetation of the siskiyou mountains, oregon and california.
\newblock {\em Ecological Monographs}, 30(3):279--338, 1960.

\bibitem{Anderson2001}
Marti~J. Anderson.
\newblock A new method for non-parametric multivariate analysis of variance.
\newblock {\em Austral Ecology}, 26(1):32--46, 2001.

\bibitem{Sanjuan2006}
Rafael Sanjuán and Santiago~F Elena.
\newblock Epistasis correlates to genomic complexity.
\newblock {\em Proceedings of the National Academy of Sciences of the United
  States of America}, 103(39):14402--14405, June 2006.

\end{thebibliography}

\newpage

\appendix

\section{Supplementary Figures}
\renewcommand\thefigure{S.\arabic{figure}}    
\setcounter{figure}{0}    

\begin{figure*}[htb]
	\centering
	\includegraphics[width=0.6\textwidth]{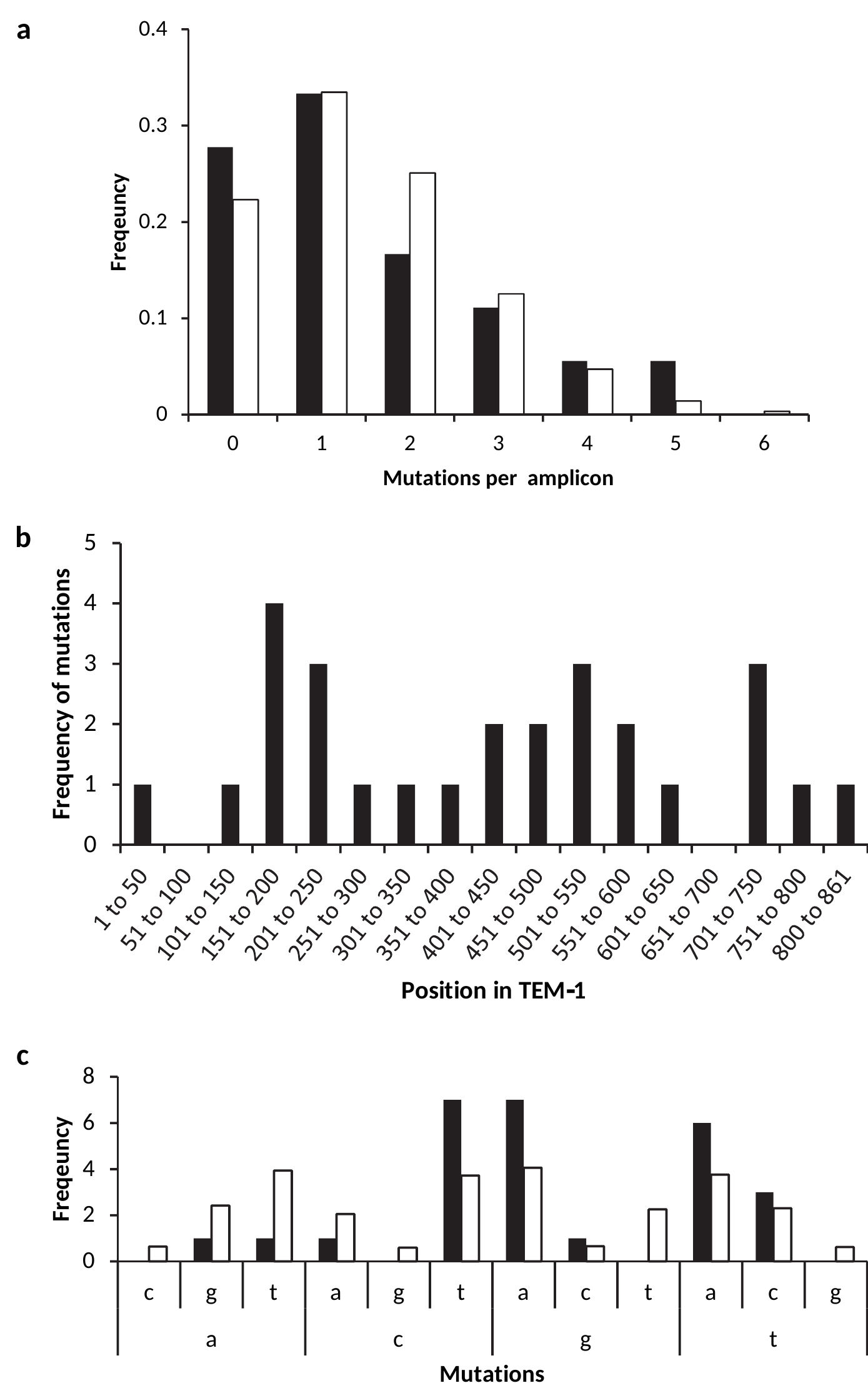}
	\caption{
		Characterisation of the TEM libraries, based on the sequencing of 18 clones. In panel a, the observed distribution of single-nucleotide mutations per allele (black bars) is shown next to the predicted Poisson distribution with a mean of 1.5 (white bars). In panel b, the number of single-nucleotide mutations is given for 50 bp bins of the TEM sequence. The 3’ end of the sequence is a 61 bp bin. Note that the mutations appear to be distributed randomly over the TEM gene. In panel c, the observed mutational spectrum for the 27 detected single-nucleotide mutations is denoted by black bars, whereas the predicted frequency is given with white bars. The prediction was generated from known bias for the polymerase and the presence of bases in TEM-1.  The lower nucleotide is the original base, and upper nucleotide the mutated base. Transitions (a <-> g, c <-> t) predominate, as well as t -> a transversions. Despite the small sample size for this analysis, the predicted and observed distributions appear to be roughly similar.
	}
	\label{fig:EPLibraries}
\end{figure*}

\begin{figure*}[t]
	\centering
	\includegraphics[width=0.9\textwidth]{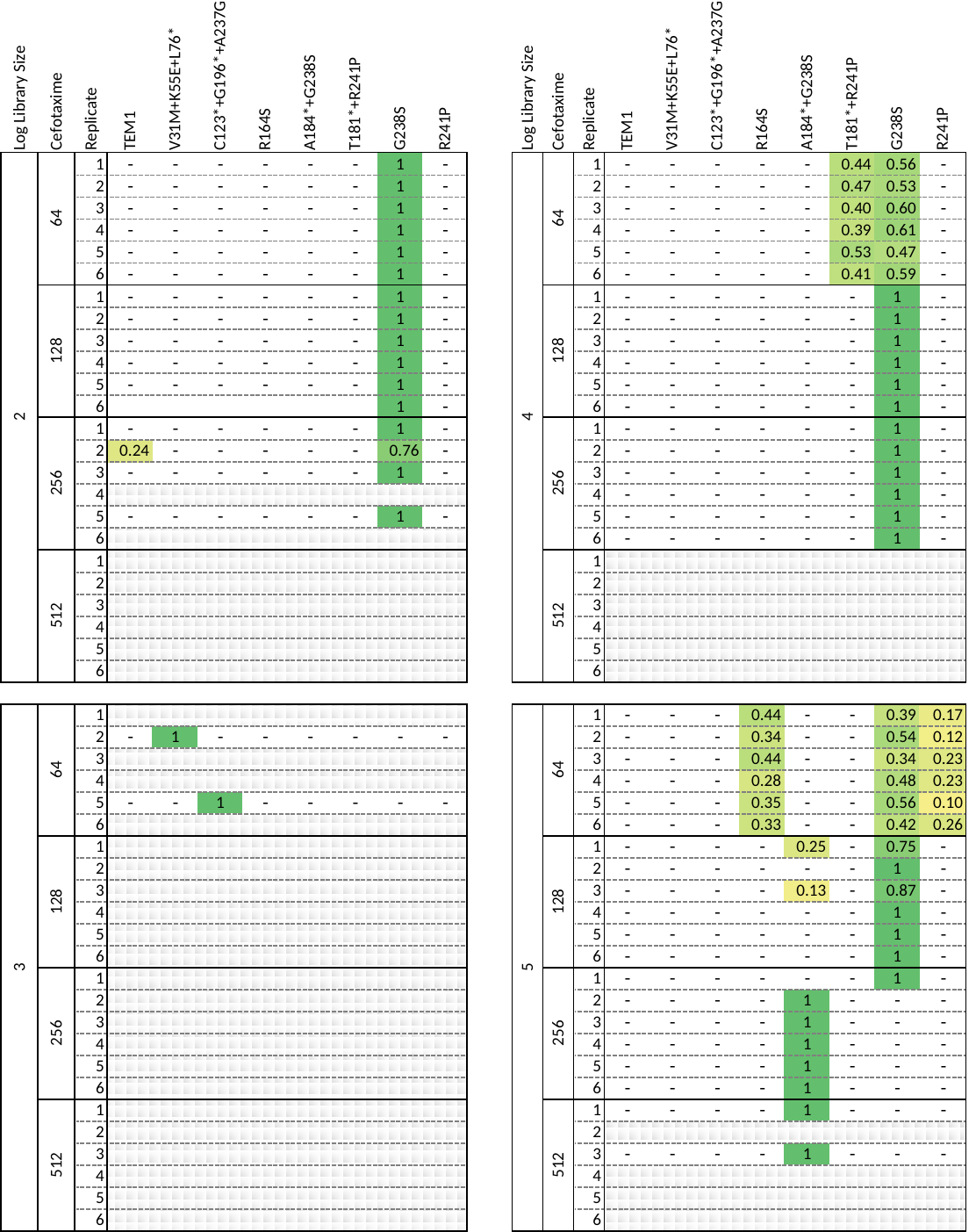}
	\caption{The haplotypes detected in populations after selection, for the libraries sampled from pool EP-PCR A. The library size, cefotaxime concentration (ng/ml) and experimental replicate are given of the left of each panel, and to the right or this the haplotype and its frequency. Grey hatches are given for populations that went extinct, and the yellow to green colors correspond with rare to fixed genotypes, respectively. Note that TEM1 is given as a haplotype, but these populations could very well be composed of an ensemble of genotypes, none of which are at a high enough frequency to be detected using our method.
	}
	\label{fig:Library1}
\end{figure*}

\begin{figure*}[t]
	\centering
	\includegraphics[width=0.9\textwidth]{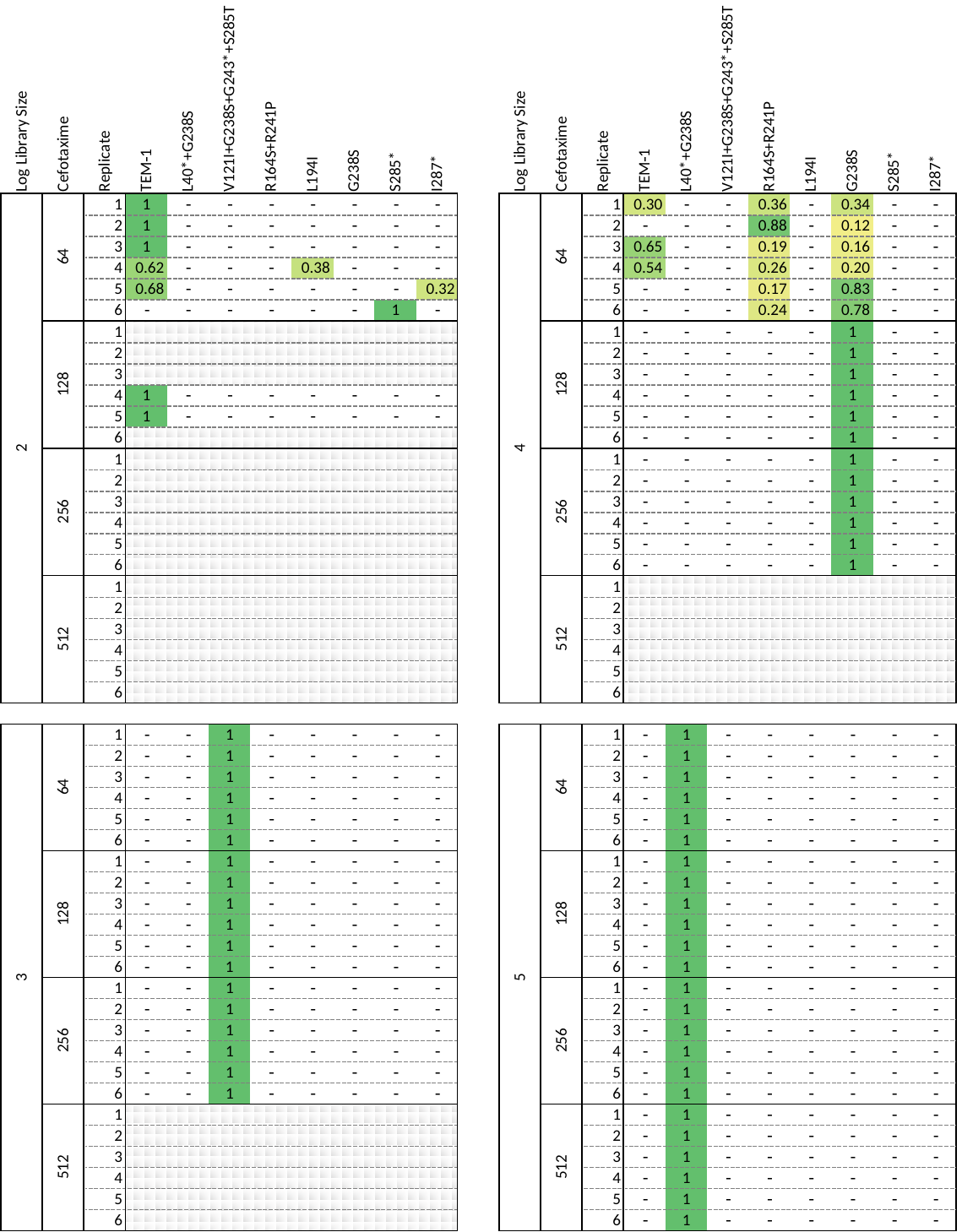}
	\caption{The haplotypes detected in populations after selection, for the libraries sampled from EP-PCR pool B. The library size, cefotaxime concentration (ng/ml) and experimental replicate are given of the left of each panel, and to the right or this the haplotype and its frequency. Grey hatches are given for populations that went extinct, and the yellow to green colors correspond with rare to fixed genotypes, respectively. Note that TEM1 is given as a haplotype, but these populations could very well be composed of an ensemble of genotypes, none of which are at a high enough frequency to be detected using our method.
	}
	\label{fig:Library2}
\end{figure*}

\begin{figure*}[t]
	\centering
	\includegraphics[width=0.9\textwidth]{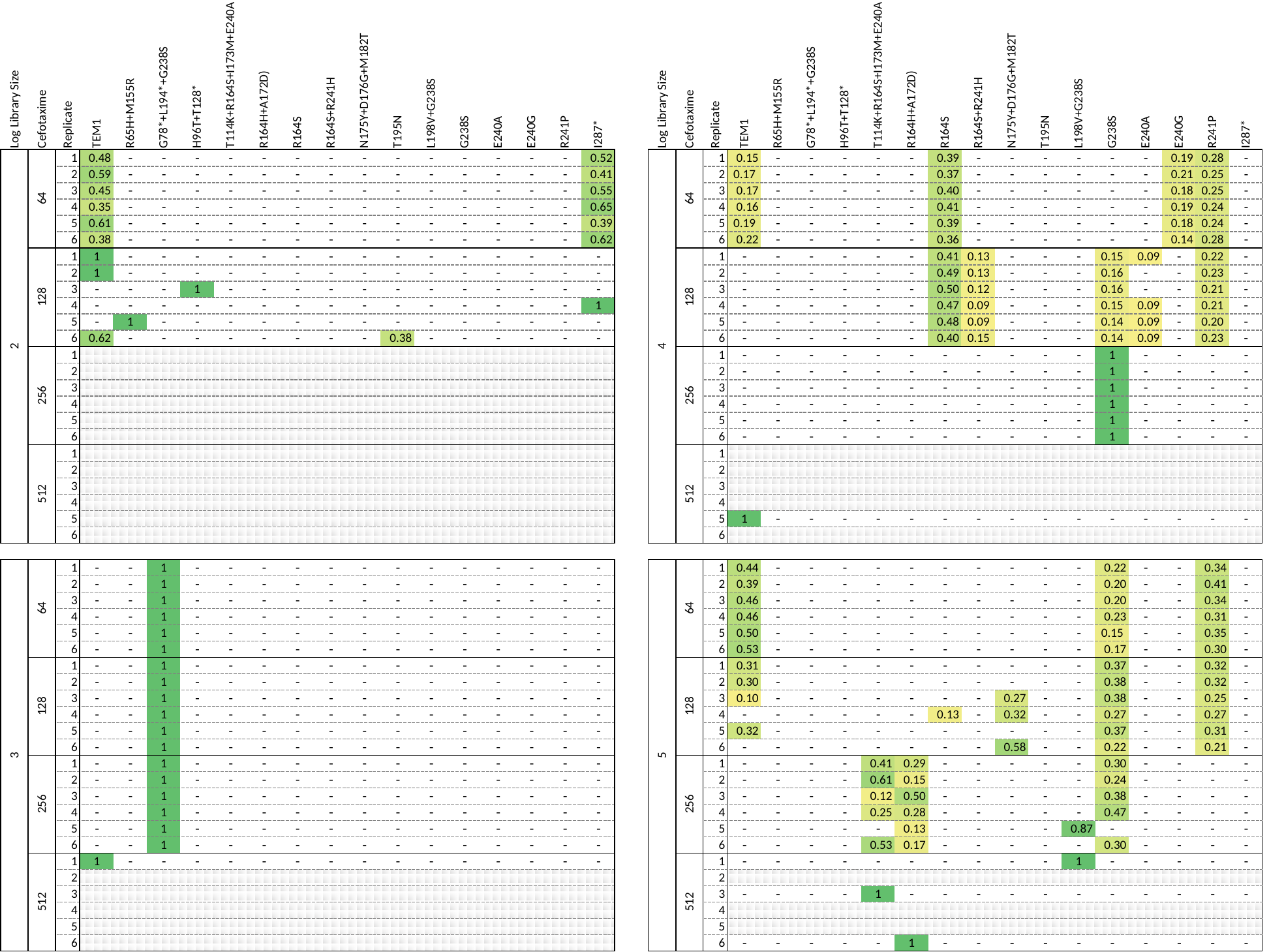}
	\caption{
		The haplotypes detected in populations after selection, for the libraries sampled from EP-PCR pool C. The library size, cefotaxime concentration (ng/ml) and experimental replicate are given of the left of each panel, and to the right or this the haplotype and its frequency. Grey hatches are given for populations that went extinct, and the yellow to green colors correspond with rare to fixed genotypes, respectively. Note that TEM1 is given as a haplotype, but these populations could very well be composed of an ensemble of genotypes, none of which are at a high enough frequency to be detected using our method.
	}
	\label{fig:Library3}
\end{figure*}

\begin{figure*}[t]
	\centering
	\includegraphics[width=0.9\textwidth]{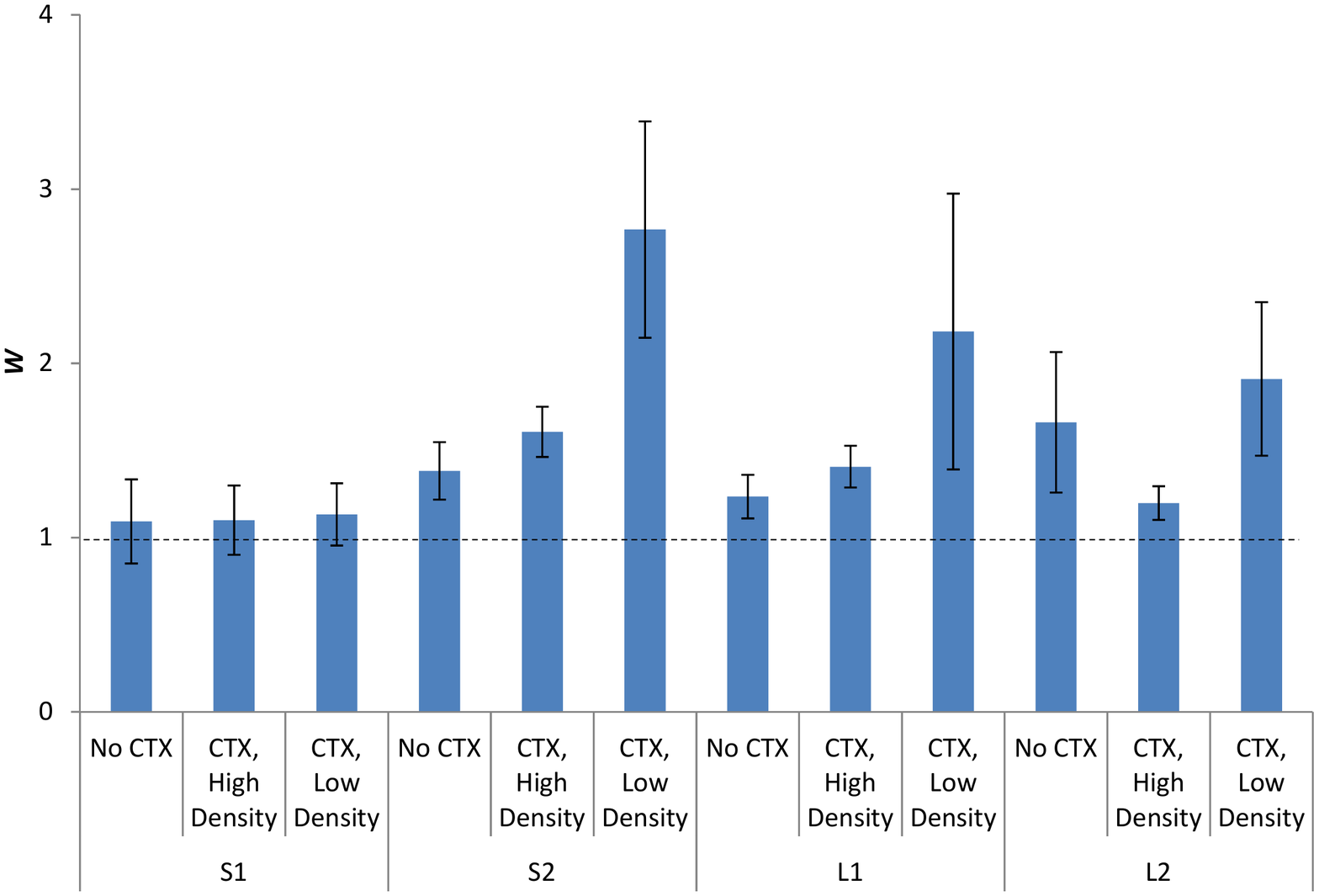}
	\caption{
		The relative fitness (\textit{W}) of different TEM alleles containing the G238S mutation are given, when in direct competition with an allele containing only the G238S mutation. Error bars represent the SEM. The mutations in alleles S1, S2, L1 and L2, and their resistance, are given in \Tref{tab:Resistance}. Competitions were performed in the absence of antibiotics and an initial density of $10^5$ cells per mL ("No CTX"), at 256 ng/mL cefotaxime and $10^5$ cells per mL ("CTX, High density"), and 256 ng/mL cefotaxime and $10^3$ cells per mL ("CTX, Low density").  The S1 allele has a similar fitness to G238S, whereas the S2 and both L alleles appear to have higher fitness, especially with cefotaxime and at a low initial cellular density.
	}
	\label{fig:Fitness}
\end{figure*}

\end{document}